\documentclass[pra,twocolumn,amsmath,amssymb,floatfix,reprint,footinbib,superscriptaddress,longbibliography,showkeys]{revtex4-2}
\usepackage{dcolumn}
\usepackage{mathtools}
\usepackage{graphicx}
\usepackage{mathrsfs}
\usepackage{mdwlist}
\usepackage{subfigure}
\usepackage{booktabs}
\usepackage{amsmath}
\usepackage{appendix}
\usepackage{extarrows}
\usepackage{dsfont}
\usepackage{amstext}
\usepackage{amssymb}
\usepackage{amsbsy}
\usepackage{bbm}
\usepackage{bm}
\usepackage{amsthm}
\usepackage{graphicx}
\usepackage{textcomp}
\usepackage{multirow}
\usepackage{threeparttable}

\usepackage{color}
\usepackage{diagbox}
\usepackage[colorlinks,citecolor=blue]{hyperref}
\definecolor{Dgreen}{RGB}{0, 100, 0}
\usepackage{url}
\usepackage[colorlinks]{hyperref}
\hypersetup{%
	plainpages=true,
	breaklinks=true, 
	hypertexnames=false, 
	pageanchor=true,
	colorlinks=true,
	linkcolor={blue},
	citecolor={blue},
	urlcolor={blue},
	anchorcolor={black}
}

\begin{document}
	\title{Generation of spin squeezing via a fully quantum degenerate parametric amplifier}
	\author{Yang Liu}
	\affiliation{Fujian Key Laboratory of Quantum Information and Quantum Optics, Fuzhou University, Fuzhou 350108, China}
	\affiliation{Department of Physics, Fuzhou University, Fuzhou 350108, China}

	\author{Jie Song}
	\affiliation{School of Physics, Harbin Institute of Technology, Harbin 150001, China}
	
	\author{Wei Qin}\thanks{wei.qin@riken.jp}
	\affiliation{Theoretical Quantum Physics Laboratory, RIKEN Cluster for Pioneering Research, Wako-shi, Saitama 351-0198, Japan}
	
	\author{Ye-Hong Chen}\thanks{yehong.chen@riken.jp}
	\affiliation{Theoretical Quantum Physics Laboratory, RIKEN Cluster for Pioneering Research, Wako-shi, Saitama 351-0198, Japan}
	\affiliation{Quantum Information Physics Theory Research Team, RIKEN Center for Quantum computing (RQC), Wako-shi, Saitama 351-0198, Japan}
	\affiliation{Fujian Key Laboratory of Quantum Information and Quantum Optics, Fuzhou University, Fuzhou 350108, China}
	\affiliation{Department of Physics, Fuzhou University, Fuzhou 350108, China}

	\author{Yan Xia}\thanks{xia-208@163.com}
	\affiliation{Fujian Key Laboratory of Quantum Information and Quantum Optics, Fuzhou University, Fuzhou 350108, China}
	\affiliation{Department of Physics, Fuzhou University, Fuzhou 350108, China}
	\begin{abstract}
		Spin squeezing is one of the most attractive methods for realizing  high-precision metrology. In this paper, we propose a protocol for generating spin squeezing in an atomic ensemble via a fully quantum degenerate parametric amplifier. We discuss the properties of generating spin squeezing with and without driving the pump cavity. Numerical simulation results show that the generated spin squeezing strength is sizable, and is able to be comparable to that obtained using a two-axis twisting (TAT) model. Moreover, we demonstrate that the protocol is experimentally feasible by introducing the corresponding experimental parameters. Therefore, the proposed protocol provides a promising approach to realize spin squeezing in photon-spin coupling systems. 
	\end{abstract}
	\maketitle
	
	\section{INTRODUCTION}\label{SEC1}
	Spin squeezing, which reduces the fluctuation noise of one quadrature in phase space but increases the fluctuation noise of the other quadrature, has shown its advantages in quantum metrology \cite{Ma2011,RevModPhys.90.035005,PhysRevLett.125.223401,PhysRevLett.127.083602}. Up to now, spin squeezing has been applied in many fields requiring high-precision measurements, such as Ramsey spectroscopy \cite{PhysRevA.46.R6797,PhysRevA.50.67,PhysRevA.64.052106,PhysRevA.81.043633}, atomic clocks \cite{PhysRevLett.117.143004,Polzik2008,PhysRevLett.104.250801}, and gravitational-wave interferometers \cite{Goda2008,Walls1981}. Due to these promising applications, significant efforts have been devoted to generating spin squeezing in many physical systems, such as molecules \cite{PhysRevLett.126.113401,PhysRevA.92.063823} and atomic ensembles \cite{Estve2008,Fadel2018,Luo2017,PhysRevA.102.023317,PhysRevLett.79.4782,PhysRevLett.83.1319,Julsgaard2001,PhysRevLett.85.1594,PhysRevLett.105.093602,Chalopin2018,PhysRevLett.122.173601,PhysRevA.101.053818,PhysRevA.102.051701,QinChenWangMiranowiczNori+2020+4853+4868,PhysRevLett.87.170402,PhysRevLett.107.013601,PhysRevLett.125.203601,PhysRevLett.118.083604}. Among the proposed protocols for atomic ensembles, the basic methods rely upon, e.g., quantum nondemolition measurement (QNDM) \cite{PhysRevLett.85.1594,PhysRevLett.105.093602}, and nonlinear one-axis twisting (OAT) \cite{Chalopin2018,PhysRevLett.122.173601} or two-axis twisting (TAT) \cite{PhysRevLett.107.013601,PhysRevLett.125.203601,PhysRevLett.118.083604} spin-spin coupling. It has been shown that different methods have different suppression effects on quantum fluctuations. For an ensemble with $N$ atoms, the maximum amounts of squeezing, obtained with QNDM, OAT, and TAT, scale as ${N^{-1/2}}$, ${N^{-2/3}}$, and ${N^{-1}}$(the ideal Heisenberg limit), respectively \cite{Ma2011}.
	
	Due to the ability to reduce the quantum fluctuation noise to the fundamental Heisenberg limit, TAT squeezing is considered as superior to other methods. Thus in the past few decades, many protocols \cite{PhysRevLett.87.170402,PhysRevLett.107.013601,PhysRevLett.125.203601,PhysRevLett.118.083604} have been proposed to generate spin squeezing by constructing the TAT interaction. These protocols include, e.g., exploiting the Raman processes in the the Bose-Einstein condensates (BEC) \cite{PhysRevLett.87.170402}, transforming from the OAT interaction \cite{PhysRevLett.107.013601}, coupling the spin ensembles with a parametric driven cavity \cite{PhysRevLett.125.203601}, and modifying a phase-locked atom-photon coupling \cite{PhysRevLett.118.083604}. However, there are always some difficulties in applying these theoretical protocols to experimental implementations. Indeed, these difficulties include, e.g., the imperfect cooling \cite{PhysRevLett.87.170402}, the imprecise time control \cite{PhysRevLett.107.013601}, the introduced squeezing noise \cite{PhysRevLett.125.203601}, and the complex pump drivings \cite{PhysRevLett.118.083604}. Recently, Macri $et\ al.$ \cite{PhysRevA.101.053818} proposed an interesting protocol for generating spin squeezing via an effective cavity-induced TAT-like interaction constructed by one-photon-two-atom excitation processes. This protocol paves a promising way to construct the TAT model and seems to be able to generate significant amount of squeezing. However, this protocol replies on a specific atom-cavity coupling (e.g., a transverse coupling), which cannot be collectively enhanced in atomic ensembles. This makes the strength of one-photon-two-atom processes extremely weak in typical ensemble-cavity systems.
	
	To address the problem, we propose a protocol for generating spin squeezing in atomic ensembles by using a fully quantum degenerate parametric amplifier (DPA) \cite{PhysRevLett.129.123602}. Here, the DPA is represented by two parametrically coupled single-mode cavities, i.e., a pump cavity and a signal cavity. An effective Hamiltonian describing the cavity-induced TAT-like interaction is obtained through tuning the system parameters. The strength of the generated spin squeezing is determined by the properties of the pump cavity, such as the initial state, driving strength, and cavity decay. Specifically, we study the generation of spin squeezing with and without driving the pump cavity mode. Numerical simulations show that with and without driving the pump cavity mode, a sizable spin squeezing can be generated. In particular, for a strong driving strength and a strong pump cavity decay, the resulting spin squeezing strength is comparable with that of the TAT squeezing. Meanwhile, we investigate the sensitivity of the generated spin squeezing to the decoherence, including the decay of the cavities, the spontaneous emission of the atoms, and the collective dephasing of the atomic ensemble. The results show that the protocol is robust to the spontaneous emission of the atoms and the decay of the signal cavity. Additionally, the experimental feasibility of the protocol is also discussed by using current experimental parameters \cite{Leghtas2015,Chang2020,Lescanne2020,Vrajitoarea2019,Hattermann2017,PhysRevLett.103.043603,PhysRevResearch.4.013207,PhysRevApplied.11.064053,Bernon2013,doi:10.1063/1.4919761,Mirhosseini2018,RevModPhys.93.025005}.
		
	The rest part of the paper is organized as follows. In Sec.~\ref{SEC2}, we give a brief description about the physical model and derive its effective Hamiltonian. In Sec.~\ref{SEC3}, we study the generation of spin squeezing in an atomic ensemble. In Sec.~\ref{SEC4}, we show the experimental feasibility of the protocol. The conclusion of the present paper is summarised in Sec.~\ref{SEC5}.
	
	\section{PHYSICAL MODEL AND EFFECTIVE DYNAMICS}\label{SEC2}
\begin{figure}
		\centering
		\includegraphics[scale=0.35]{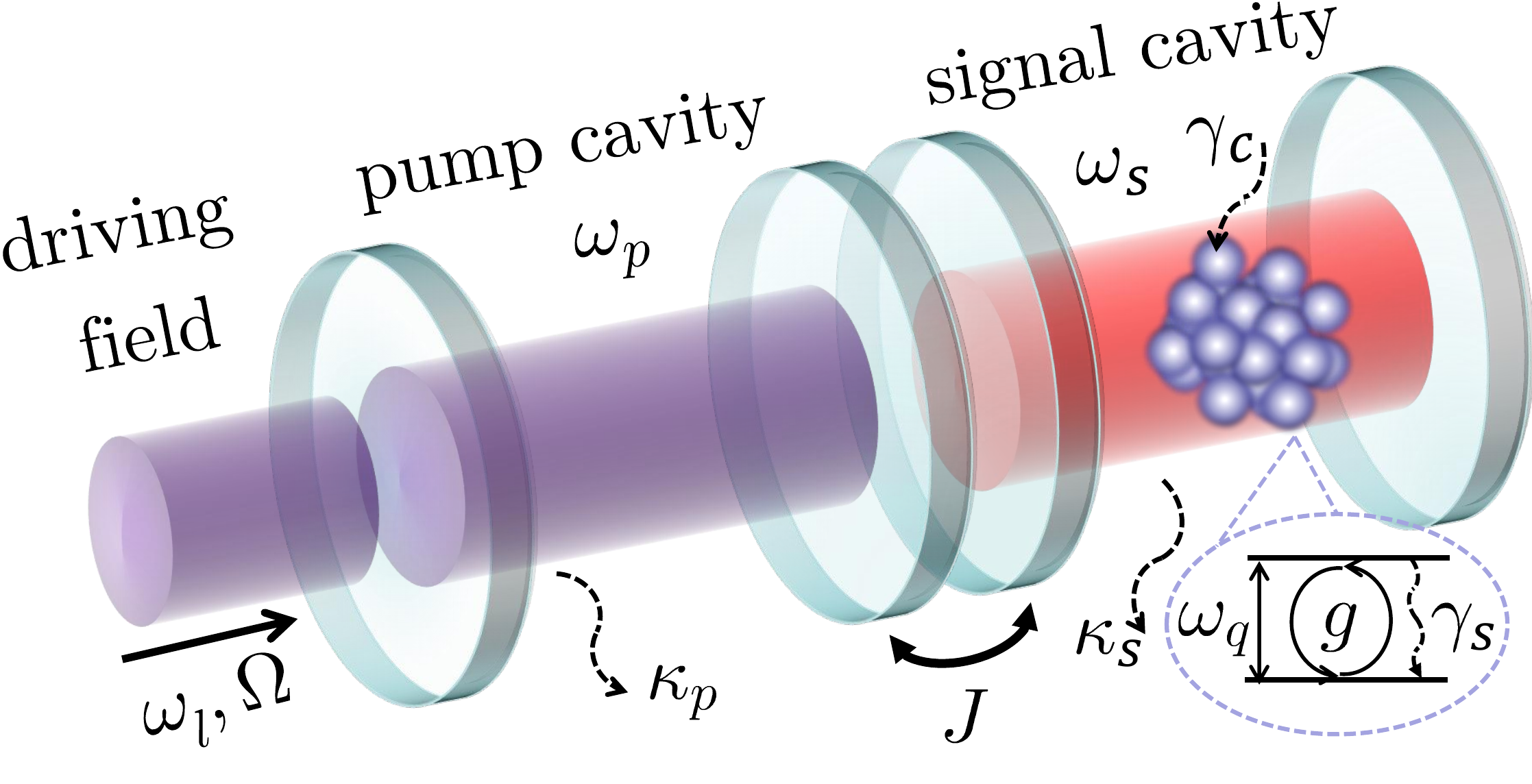}
		\caption[Fig1]{Physical model of the protocol. Two single-mode cavities, the pump cavity and the signal cavity, are coupled with a parametric coupling of strength $J$. An ensemble of $N$ identical two-level atoms is located inside the signal cavity, and is coupled to this cavity with a single-photon single-atom coupling strength $g$. $\omega_{p}(\omega_{s})$ is the resonance frequency of the pump (signal) cavity. A driving field of frequency $\omega_l$ and amplitude $\Omega$ is applied to the pump cavity. $\omega_q$ is the transition frequency of atoms, $\kappa_{p}(\kappa_{s})$ is the single-photon dissipation rate of the pump (signal) cavity, $\gamma_s$ is the spontaneous emission rate of atoms, and $\gamma_c$ represents the collective dephasing of the atomic ensemble.}
		\label{fig1}
	\end{figure}

	In this paper, as shown in Fig.~\ref{fig1}, we consider a system consisting of two parametrically coupled single-mode cavities (a pump cavity and a signal cavity), and an ensemble of $N$ identical two-level atoms placed in the signal cavity. The resonance frequencies of the pump cavity and the signal cavity are assumed to be $\omega_p$ and $\omega_s$, respectively. The parametric coupling of strength $J$ describes a nonlinear conversion between a single pump photon and a pair of signal photons. Note that the strength $J$, which ranges from some tens of kHz to some tens of MHz, has been realized in recent experimental advances \cite{Leghtas2015,Chang2020,Lescanne2020,Vrajitoarea2019}. Therefore, the pump cavity and the signal cavity constitute a fully quantum degenerate parametric amplifier (DPA). We apply a classical driving field of frequency $\omega_l$ and amplitude $\Omega$ to the pump cavity. Meanwhile, the atoms of transition frequency $\omega_q$ couple to the signal cavity with a strength $g$. The Hamiltonian of the system can accordingly be written as (in the units of $\hbar$) \cite{PhysRevLett.127.093602}
	\begin{eqnarray}\label{sec2eq1}
	\hat{H}_\text{sys}&=&\omega_s\hat{a}^\dagger_s\hat{a}_s+\omega_p\hat{a}^\dagger_p\hat{a}_p+\omega_q\hat{S}_z+g(\hat{a}_s\hat{S}_++\hat{a}^\dagger_s\hat{S}_-)
	\cr\cr
	&&+\left(\frac{\Omega^*}{2}\hat{a}_pe^{i\omega_lt}+\frac{\Omega}{2}\hat{a}^\dagger_pe^{-i\omega_lt}\right)+J(\hat{a}_p\hat{a}^{\dagger2}_s+\hat{a}^\dagger_p\hat{a}^2_s),\cr\cr&&
	\end{eqnarray}
	where $\hat{a}^\dagger_{s}(\hat{a}^\dagger_{p})$ and $\hat{a}_{s}(\hat{a}_{p})$ are the creation operator and the annihilation operator of the signal (pump) cavity, respectively. Moreover, ${\hat{S}_u=\sum_k\hat{\sigma}^u_k/2}$ $(k=1,2,\dots,N)$ is the collective spin operator and $\hat{\sigma}^u_k$ ${(u=x,y,z)}$ is the Pauli operator of the $k$th atom. ${\hat{S}_{+}=\hat{S}_x+i\hat{S}_y}$ and ${\hat{S}_{-}=\hat{S}_x-i\hat{S}_y}$ are the raising and lowering operators of the collective spin, respectively. For simplicity, we set hereafter ${g_c=\sqrt{N}g}$ as the strength of the collective coupling between the ensemble and the cavity, and further assume that $\omega_{p}\simeq\omega_{l}\simeq2\omega_{q}\simeq2\omega_{s}$. In the rotation frame of reference of  $\hat{H}_0=\omega_l(\hat{a}^\dagger_s\hat{a}_s+2\hat{a}^\dagger_p\hat{a}_p+\hat{S}_z)$, when ${\omega_l\gg\{g_c,\,J\}}$, one can obtain
	\begin{eqnarray}\label{sec2eq2}
		\hat{H}_I&=&\delta_p\hat{a}^\dagger_p\hat{a}_p+\delta_s\hat{a}^\dagger_s\hat{a}_s+\delta_q\hat{S}_z\cr\cr
		&&+(J\hat{a}_p\hat{a}^{\dagger 2}_s+g\hat{a}_s\hat{S}_++\frac{\Omega^*}{2}\hat{a}_p+\text{H.c.}),
	\end{eqnarray}
	where ${\delta_{q}=\omega_{q}-\omega_l}$, ${\delta_{s}=\omega_{s}-\omega_l}$, and ${\delta_p=\omega_p-2\omega_l}$. 
	
	When the pump cavity suffers a cavity decay, the dynamics of the system can be estimated by the master equation in the Lindblad form \cite{QIC}
	\begin{eqnarray}\label{sec2eq3}
		\dot{\rho}=i[\rho,\hat{H}_I]+\kappa_{p}\mathcal{L}(\hat{a}_p)\hat{\rho},
	\end{eqnarray}
	where for an arbitrary operator $\hat{o}$, the standard Lindblad superoperator is defined as
	\begin{eqnarray}\label{sec2eq4}
		\mathcal{L}(\hat{o})\rho=\hat{o}\rho\hat{o}^\dagger-\frac{1}{2}(\hat{o}^\dagger\hat{o}\rho+\rho\hat{o}^\dagger\hat{o}).
	\end{eqnarray}
	Here, $\kappa_{p}$ is the single-photon dissipation rate (i.e., the cavity decay rate) of the pump cavity. Due to the presence of the driving $\Omega$ of the pump mode, we first set ${\hat{c}_p=\hat{a}_p+d}$ to derive the effective Hamiltonian of the system. Thereafter, the Hamiltonian in Eq.~(\ref{sec2eq2}) is transformed to
	\begin{eqnarray}\label{sec2eq5} 
		\hat{H}_h&=&\delta_p(\hat{c}^{\dagger}_p-d^*)(\hat{c}_p-d)+\delta_s\hat{a}^\dagger_s\hat{a}_s+\delta_q\hat{S}_z\cr\cr
		&&+\left[J(\hat{c}_p-d)\hat{a}^{\dagger2}_s+g\hat{a}_s\hat{S}_++\frac{\Omega^*}{2}\hat{c}_p+\text{H.c.}\right].
	\end{eqnarray} 
	Meanwhile, the second term of the master equation in Eq.~(\ref{sec2eq3}) becomes
	\begin{eqnarray}\label{sec2eq6} 
		\kappa_p\mathcal{L}(\hat{a}_p)\rho&=&\kappa_p\mathcal{L}(\hat{c}_p)\rho-i[\hat{H}_{\text{me}},\rho], 
	\end{eqnarray}
	where ${\hat{H}_{\text{me}}=-i\kappa_p(d^*\hat{c}_p-d\hat{c}^\dagger_p)/2}$. Then, the total Hamiltonian can be described as
	\begin{eqnarray}\label{sec2eq7}
		\hat{H}_\text{tot}&=&\hat{H}_h+\hat{H}_{\text{me}}\cr\cr
		&=&\delta_p\hat{c}^{\dagger}_p\hat{c}_p+\delta_s\hat{a}^\dagger_s\hat{a}_s+\delta_q\hat{S}_z{-J}\left(d\hat{a}^{\dagger2}_s+d^*\hat{a}^{2}_s\right)\cr\cr
		&&+\left(\Omega_d\hat{c}_p+J\hat{c}_p\hat{a}^{\dagger2}_s+g\hat{a}_s\hat{S}_++\text{H.c.}\right),
	\end{eqnarray} 
	and the dynamics of the system can be modelled by the master equation 
	\begin{eqnarray}\label{sec2eq8} 
		\dot{\rho}=-i[\hat{H}_{\text{tot}},\rho]+\kappa_p\mathcal{L}(\hat{c}_p)\rho,
	\end{eqnarray} 
	where $\Omega_d=(\Omega^* -2d^*\delta_p-i\kappa_p d^*)/2$. For simplicity, we have set ${d=\Omega/(2\delta_p-i\kappa_{p})}$ to eliminate the driving term with $\Omega_d$.
	
	According to the second and fourth terms in Eq.~(\ref{sec2eq7}), the signal cavity is squeezed \cite{Ran2020AUG5,Liu2020JUN,Kang2022AUG15,Wang2019FEB19,Wang2019OCT,Chen2022OCT}. Applying the Bogoliubov transformation \cite{QinChenWangMiranowiczNori+2020+4853+4868,PhysRevLett.126.023602,PhysRevA.100.062501,Wang2019FEB19,Wang2019OCT,Chen2022OCT}
	$$\hat{a}_{\texttt{S}}=\cosh(r)\hat{a}_s+\sinh(r)e^{i\theta_{\texttt{S}}}\hat{a}^{\dagger}_s,$$
	one can diagonalize the signal-cavity Hamiltonian in Eq.~(\ref{sec2eq7}), and obtain the Hamiltonian
	\begin{eqnarray}\label{sec2eq9}
		\hat{\mathcal{H}}&=&\delta_p\hat{c}^{\dagger}_p\hat{c}_p+\delta_{\texttt{S}}\hat{a}^\dagger_{\texttt{S}}\hat{a}_{\texttt{S}}+\delta_q\hat{S}_z\cr\cr		&&+\Big\{J\big[\cosh(r)\hat{a}_{\texttt{S}}+\sinh(r)e^{i\theta_\texttt{S}}\hat{a}^\dagger_{\texttt{S}}\big]^2\hat{c}^\dagger_p\cr\cr &&+g\big[\cosh(r)\hat{a}_{\texttt{S}}+\sinh(r)e^{i\theta_\texttt{S}}\hat{a}^\dagger_{\texttt{S}}\big]\hat{S}_++\text{H.c.}\Big\},
	\end{eqnarray}
	where ${\delta_{\texttt{S}}=\sqrt{\delta^2_s-(2J\vert d\vert)^2}}$,  $\theta_\texttt{S}=-\arctan(\text{Im}(d)/\text{Re}(d))$, and $r=(1/4)\ln[(\delta_s+2J\vert d\vert)/(\delta_s-2J\vert d\vert)]$. Here, we set ${\delta_s\gg2J\vert d\vert}$, which makes the cavity squeezing parameter $r$ approach $0$. Then, we have ${\cosh(r)\rightarrow1}$, ${\sinh(r)\rightarrow r}$, and ${\sinh(2r)\rightarrow 2r}$. Thereafter, the Hamiltonian $\hat{\mathcal{H}}$ in Eq.~(\ref{sec2eq9}) becomes approximated by
	\begin{eqnarray}\label{sec2eq10} 
		\hat{\mathcal{H}}&\simeq&\delta_{\texttt{S}}\hat{a}^\dagger_{\texttt{S}}\hat{a}_{\mathtt{S}}+\delta_q\hat{S}_z+\delta_p\hat{c}^{\dagger}_p\hat{c}_p\cr\cr
		&&+\Big[J(\hat{a}_{\texttt{S}}+re^{i\theta_\texttt{S}}\hat{a}^\dagger_{\texttt{S}})^2\hat{c}^\dagger_p+g(\hat{a}_{\texttt{S}}+re^{i\theta_\texttt{S}}\hat{a}^\dagger_{\texttt{S}})\hat{S}_++\text{H.c.}\Big]. \cr\cr&&
	\end{eqnarray} 
	Furthermore, when the condition ${\delta_\texttt{S}\gg\{J,\,g_c\}}$ is satisfied, according to the works in Refs.~\cite{p07-060,PhysRevA.95.032124}, the dynamics of the system can, up to the third order, be described by the effective Hamiltonian
	\begin{eqnarray}\label{sec2eq12}
		\hat{\mathcal{H}}_\text{eff}&=&\hat{\mathcal{H}}^{(1)}_\text{eff}+\hat{\mathcal{H}}^{(2)}_\text{eff}+\hat{\mathcal{H}}^{(3)}_\text{eff},\cr\cr
		\hat{\mathcal{H}}^{(1)}_\text{eff}&=&\delta_p\hat{c}^\dagger_p\hat{c}_p+\delta_q\hat{S}_z+Jr(2\hat{a}^{\dagger}_{\texttt{S}}\hat{a}_{\texttt{S}}+1)(e^{-i\theta_{\texttt{S}}}\hat{c}_p+\text{H.c.}),\cr\cr
		\hat{\mathcal{H}}^{(2)}_\text{eff}&=&-\frac{J^2}{2\delta_\texttt{S}}(2\hat{c}^\dagger_p\hat{c}_p+4\hat{c}^\dagger_p\hat{c}_p\hat{a}^\dagger_\texttt{S}\hat{a}_\texttt{S}-\hat{a}^\dagger_\texttt{S}\hat{a}^\dagger_\texttt{S}\hat{a}_\texttt{S}\hat{a}_\texttt{S})\cr\cr
		&&-\frac{g^2}{\delta_\texttt{S}}(2\hat{a}^\dagger_\texttt{S}\hat{a}_\texttt{S}\hat{S}_z+\hat{S}_+\hat{S}_-),\cr\cr
		\hat{\mathcal{H}}^{(3)}_\text{eff}&=&\frac{g^2}{\delta_\texttt{S}}\left[\left(\frac{J}{\delta_{\texttt{S}}}\hat{c}_p+re^{i\theta_{\texttt{S}}}\right)\hat{S}_+^2+\text{H.c.}\right],
	\end{eqnarray}
	where $\hat{\mathcal{H}}^{(\mathcal{J})}_\text{eff}$ (${\mathcal{J}=1,2,3}$) represents the $\mathcal{J}$th order processes of the effective Hamiltonian. Here, we also have neglected the fast-oscillating terms by the rotating wave approximation. Note that the Hamiltonian $\hat{\mathcal{H}}^{(3)}_\text{eff}$ in Eq.~(\ref{sec2eq12}) contains the nonlinear interactions of the atomic ensembles, which is crucial for generating spin squeezing in the protocol. However, these nonlinear interactions are  third-order processes whose effect would be masked by that of the lower-order interactions. 
	
	To eliminate the influence of the lower-order interactions, we tune the parameters and make some proper assumptions as shown in the following. First, we assume that the signal cavity is well cooled so that the signal cavity is in the vacuum state, i.e., ${\langle\hat{a}^\dagger_\texttt{S}\hat{a}_\texttt{S}\rangle\simeq0}$. Therefore, all the interactions involving $\hat{a}^\dagger_\texttt{S}\hat{a}_\texttt{S}$, like the terms $\hat{a}^\dagger_p\hat{a}_p\hat{a}^\dagger_\texttt{S}\hat{a}_\texttt{S}$ and $\hat{a}^\dagger_\texttt{S}\hat{a}_\texttt{S}\hat{S}_z$, can be neglected. Meanwhile, ${\delta_p=J^2/\delta_\texttt{S}}$ is set to eliminate the Stark shift of the pump cavity. Furthermore, we assume that in the ensemble, the number of excited atoms is much smaller than the total number of atoms. Under this condition, the spin operator $\hat{S}_z$ can be divided into two parts, i.e.,
	\begin{eqnarray}\label{sec2eq13}
		\hat{S}_z=-\frac{N}{2}+\Delta\hat{S}_z,
	\end{eqnarray}
	where $\Delta\hat{S}_z$ is a small fluctuation around the ground state. After substituting Eq.~(\ref{sec2eq13}) into the relation $\hat{S}_z^2-\hat{S}_z+\hat{S}_+\hat{S}_-=N(N+1)/4$, we obtain (ignoring the constant terms)
	\begin{eqnarray}\label{sec2eq14}
		\hat{S}_+\hat{S}_-\simeq (N+1)\Delta S_z=(N+1)\hat{S}_z.
	\end{eqnarray}
	This means that, when the atomic ensemble is in the low-excitation regime, the effect of the coupling $\hat{S}_+\hat{S}_-$ is to produce a Stark shift given by $(N+1)\hat{S}_z$. Therefore, one can choose ${\delta_q=(N+1)g^2/\delta_\texttt{S}}$ to eliminate he influence of the term $\hat{S}_+\hat{S}_-$ effectively. Meanwhile, the effect of this term can be eliminated completely by introducing an auxiliary atomic level and a driven optical cavity \cite{PhysRevLett.113.203601,Lauk_2020,PhysRevA.103.023706} (see more details in Appendix A). In the rest of paper and numerical simulations, we assume the effect of $\hat{S}_+\hat{S}_-$ has been fully compensated and $\delta_{q}=0$. Correspondingly, the effective Hamiltonian of the system becomes
	\begin{eqnarray}\label{sec2eq15}
		\hat{\mathcal{H}}_{\text{eff}}\approx g_{\text{eff}}(\hat{c}_p-d)\hat{S}^2_++\frac{J^2}{\delta_\mathtt{S}}d^*\hat{c}_p+\text{H.c.},
	\end{eqnarray}
	where we have made the approximation ${r\rightarrow J\vert d\vert/\delta_{\texttt{S}}}$ and assumed ${g_\text{eff}=g^2J/{\delta_\mathtt{S}^2}}$. Then, by taking corresponding reverse transformations, one can convert the effective Hamiltonian in Eq.~(\ref{sec2eq15}) back to the original frame and obtain
	\begin{eqnarray}\label{sec2eq16}
		\hat{H}_{\text{eff}}=g_\text{eff}(\hat{a}_p\hat{S}^2_++\hat{a}^\dagger_p\hat{S}^2_-)+\left(\frac{\Omega^*}{2}\hat{a}_p+\frac{\Omega}{2}\hat{a}^\dagger_p\right).
	\end{eqnarray} 	
	Meanwhile, the effective Lindblad-type master equation in Eq.~(\ref{sec2eq8}) becomes 
	\begin{eqnarray}\label{sec2eq17}
		\dot{\rho}=i[\rho,\hat{H}_{\text{eff}}]+\kappa_p\mathcal{L}(\hat{a}_p)\rho.
	\end{eqnarray} 
	The first term in Eq.~(\ref{sec2eq16}) indicates that the system involves a cavity-induced TAT-like interaction which can be used to generate spin squeezing \cite{PhysRevA.101.053818}. The built-in mechanism is that when the pump cavity is in a coherent state with ${\langle\hat{a}_p\rangle=\beta}$, the effective Hamiltonian  in Eq.~(\ref{sec2eq16}) can be approximated as
	\begin{eqnarray}\label{sec2eq17a}
		\hat{H}'_{\text{eff}}=g_\text{eff}(\beta\hat{S}^2_++\beta^*\hat{S}^2_-),
	\end{eqnarray}
	i.e., the TAT interaction. Moreover, the TAT model in Eq.~(\ref{sec2eq17a}) can be established through keeping the pump cavity in a state with $\langle\hat{a}_p\rangle\not=0$ during the evolution. According to Eqs.~(\ref{sec2eq16}) and (\ref{sec2eq17}), keeping the state in the pump cavity unchanged can be achieved effectively by increasing the ratio between $\Omega$ and $g_\text{eff}$. The detail is that for a larger ratio $\Omega/g_\text{eff}$, the pump cavity stays in a coherent state with amplitude $d_0=i\Omega/\kappa_{p}$ efficiently for a longer period (see below for numerical demonstrations). For convenience, we take the coherent state as the quasi-steady state of the pump cavity in the remaining of the text.
	
	So far, we have considered a model, which contains only the photon loss of the pump mode. In reality, there always exists some dissipative processes for the atomic ensemble and the signal cavity. Here, we assume that the system suffers from dissipation induced by the ensemble collective dephasing, the atomic spontaneous emission, and the signal cavity single-photon loss. The dynamics of the system can therefore be described by the following master equation
	\begin{eqnarray}\label{sec2eq18}
		\dot{\rho}_I{=}i[\rho_I,\hat{H}_I]{+}\!\!\left[\sum_{v=s,p}\kappa_{v}\mathcal{L}(\hat{a}_v){+}\sum_{k=1}^{N}\gamma_{s}\mathcal{L}(\hat{\sigma}^-_k){+}\gamma_{c}\mathcal{L}(\hat{S}_z)\right]\!\rho_I,\cr\cr
	\end{eqnarray}
	where $\rho_I$ is the density operator under the full dynamics of the system. The parameters $\kappa_{s}$, $\gamma_s$, and $\gamma_c$ are the single-photon loss rate of the signal cavity, the spontaneous emission rate of atoms, and the collective dephasing rate of the atomic ensemble, respectively. In general, the third term in Eq.~(\ref{sec2eq18}), i.e.,  $\sum_{k=1}^{N}\gamma_s\mathcal{L}(\hat{\sigma}^-_k)\rho_I$, makes numeral simulations of a large-size ensemble dynamics extremely difficult because the demanded computation resources increase exponentially with the total number of atoms $N$. However, the system dynamics involves only the zero momentum mode of the atomic ensemble, and also does not mix it with other nonzero momentum modes. Thus according to Refs.~\cite{PhysRevA.101.053818,PhysRevA.98.063815,PhysRevA.95.063824,QinChenWangMiranowiczNori+2020+4853+4868}, the third term in Eq.~(\ref{sec2eq18}) can be reduced to
	\begin{eqnarray}\label{sec2eq21}
		\sum_{k}\mathcal{L}(\hat{\sigma}^-_k)\rho_I=\frac{1}{N}\mathcal{L}(\hat{S}_-)\rho_I.
	\end{eqnarray}
	Then, the full master equation in Eq.~(\ref{sec2eq18}) becomes
	\begin{eqnarray}\label{sec2eq22}
		\dot{\rho}_I=i[\rho_I,\hat{H}_I]+\!\!\left[\sum_{v=p,s}\!\!\kappa_{v}\mathcal{L}(\hat{a}_{v})+\frac{\gamma_s}{N}\mathcal{L}(\hat{S}_-)+\gamma_{c}\mathcal{L}(\hat{S}_z)\right]\!\rho_I.\cr\cr
	\end{eqnarray}
	Accordingly, the effective master equation in Eq.~(\ref{sec2eq17}) is transformed to
	\begin{eqnarray}\label{sec2eq23}
		\dot{\rho}_{\text{eff}}&=&i[\rho_{\text{eff}},\hat{H}_{\text{eff}}]\cr\cr
		&&+\!\left[\sum_{v=s,p}\kappa_{v}\mathcal{L}(\hat{a}_v){+}\frac{\gamma_s}{N}\mathcal{L}(\hat{S}_-){+}\gamma_{c}\mathcal{L}(\hat{S}_z)\right]\rho_{\text{eff}},
	\end{eqnarray} 
	where $\rho_{\text{eff}}$ represents the density operator under the effective dynamics of the system. Note that, the part $\kappa_{s}\mathcal{L}(\hat{a}_s)$ in Eq.~(\ref{sec2eq17}) can be subtracted since the signal cavity has been decoupled from the dynamics of the effective Hamiltonian in Eq.~(\ref{sec2eq16}).
	
	As demonstrated above, we have obtained the cavity-induced TAT-like interaction effectively, and the Lindblad-type master equation to simulate the dynamics of the system. These enable us to generate spin squeezing and study the properties of the generated spin squeezing under different decoherence noise sources.
	
	\section{GENERATING SPIN SQUEEZING}\label{SEC3}
	
	In this section, we investigate the generation of spin squeezing through theoretical analyses and numerical simulations. First, we need to introduce the spin squeezing parameter $\xi^2_R$ proposed by Wineland $et\ al.$ \cite{PhysRevA.46.R6797,PhysRevA.50.67,Ma2011}
	\begin{eqnarray}\label{sec3eq1}
		\xi^2_R=N\frac{\langle(\hat{\boldsymbol{S}}\cdot\boldsymbol{n}_\bot)^2\rangle-\langle\hat{\boldsymbol{S}}\cdot\boldsymbol{n}_\bot\rangle^2}{\vert\langle\hat{\boldsymbol{S}}\rangle\vert^2}, 
	\end{eqnarray}
	where $\hat{\boldsymbol{S}}=\hat{S}_x\boldsymbol{e}_x+\hat{S}_y\boldsymbol{e}_y+\hat{S}_z\boldsymbol{e}_z$ and $\boldsymbol{e}_{u}(u=x,y,z)$ is the unit vector along the $u$ direction. The unit vector $\boldsymbol{n}_\bot$ is along the direction minimizing the numerator. The spin squeezing parameter represents the ratio of the fluctuations between a quantum state of interest and a coherent spin state (CSS) in Ramsey spectroscopy. Here, the CSS acts as a noise-reference state. It is seen from Eq.~(\ref{sec3eq1}) that, when $\xi^2_R<1$, the phase sensitivity of the state of interest is improved over the standard quantum limit (SQL), i.e., this state is squeezed. Note that, a smaller $\xi_R^2$ indicates a stronger spin squeezing. In this paper, we choose the parameter $\xi^2_R$ to characterize the strength of the generated spin squeezing. 
	
	To numerically simulate the evolution of the system, we use the Monte Carlo approach (i.e., the quantum-jump method) \cite{PhysRevLett.68.580}, where individual quantum trajectories of the system evolve under a non-Hermitian Hamiltonian, and then are randomly interrupted by quantum jumps. Moreover, the dynamics of the system is regarded as an ensemble average over these trajectories of the system wave functions. The number of the involved trajectories is larger and the description on the dynamics of the system is more precise. However, simulating a large number of trajectories also requires many computation resources. Therefore, after balancing the precise of the numerical simulation and the requirement of the computation resources, we take the average over $1000$ trajectories to calculate the dynamics of the system. 
	
	In the following two subsections, we assume that initially, the pump cavity is in a coherent state $\vert\alpha\rangle_p$, where ${\alpha=\vert\alpha\vert e^{i\varphi}}$ is the complex amplitude with an argument ${\varphi=\arg(\alpha)}$. At the same time, the signal cavity and the atomic ensemble are well cooled to their ground states, i.e., $\vert0\rangle_s$ and $\vert l,-l\rangle_e$, respectively, where ${l=N/2}$. Here, $\vert j,m_z\rangle_e$ represents a collective spin state of an ensemble of $N$ spin-1/2 atoms, with an orbital angular momentum quantum number and a magnetic quantum number ${m_z\in\{-l,-l+1,\dots,l-1,l\}}$.
	
	\subsection{WITH DRIVING THE PUMP CAVITY}\label{SEC3a}
	In this subsection, we study the generation of spin squeezing in the case where the pump cavity is driven by external fields and is initialized in the quasi-steady coherent state, i.e., $\alpha=d_0=i\Omega/\kappa_{p}$.
	
	\begin{figure}
		\subfigure{\includegraphics[scale=0.56]{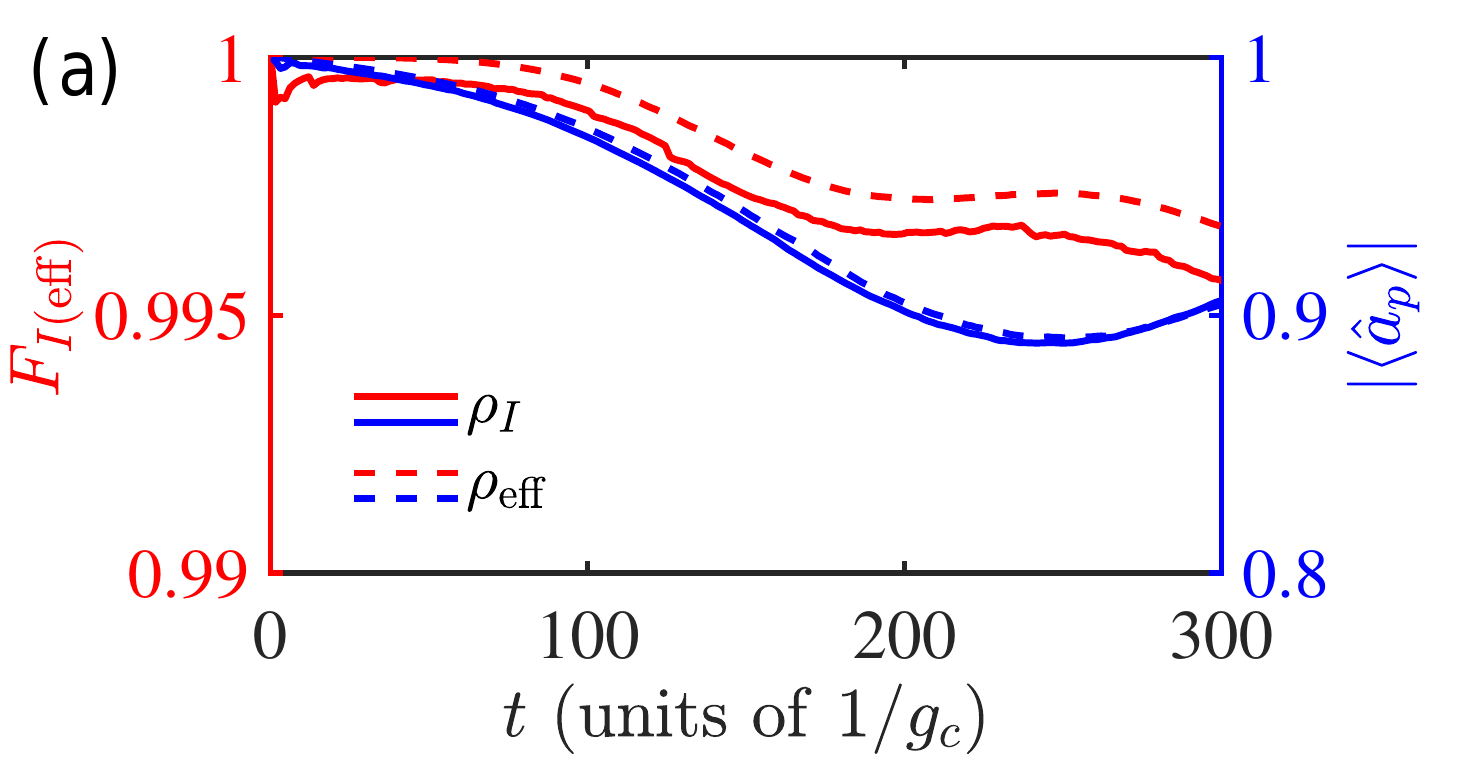}\label{fig2a}}
		\subfigure{\includegraphics[scale=0.55]{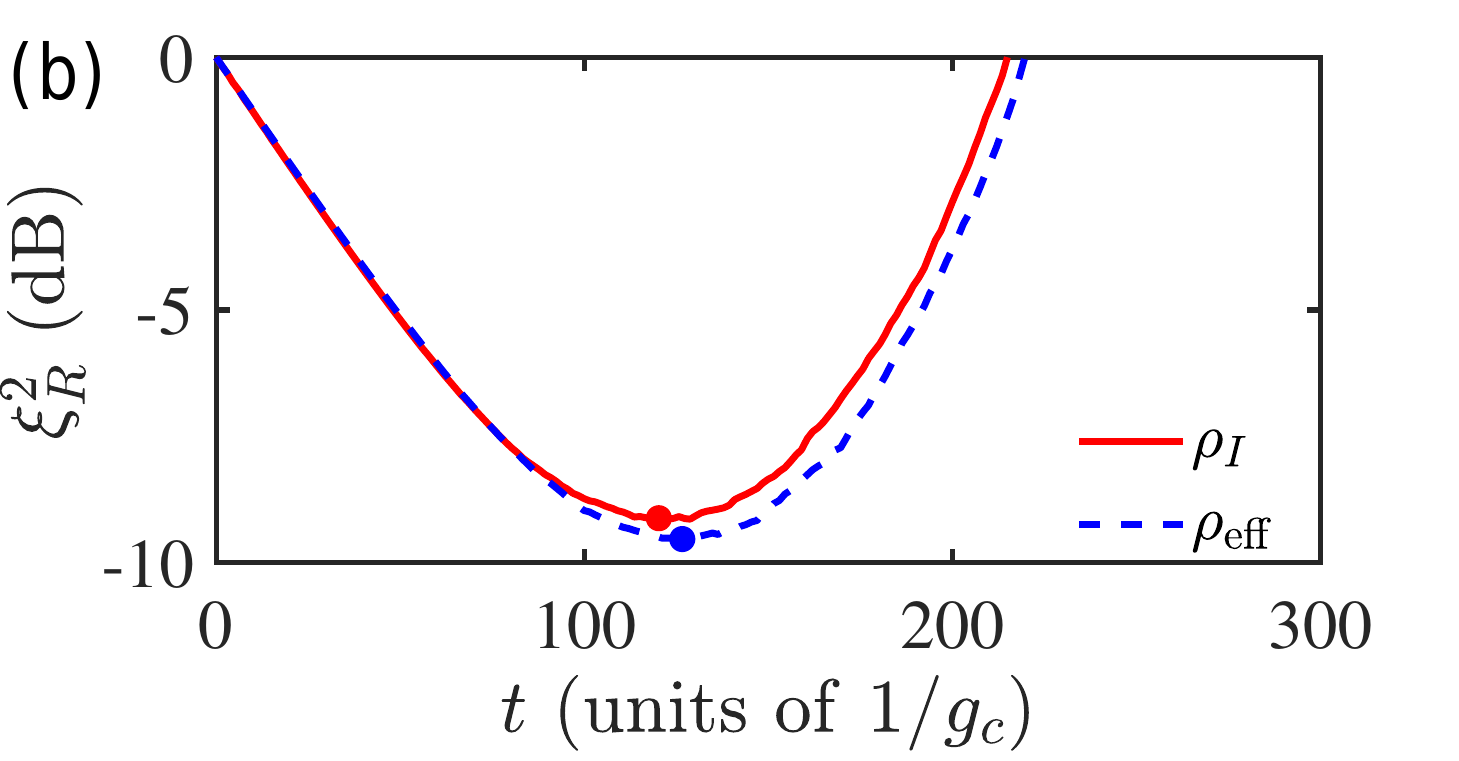}\label{fig2b}}
		\subfigure{\includegraphics[scale=0.40]{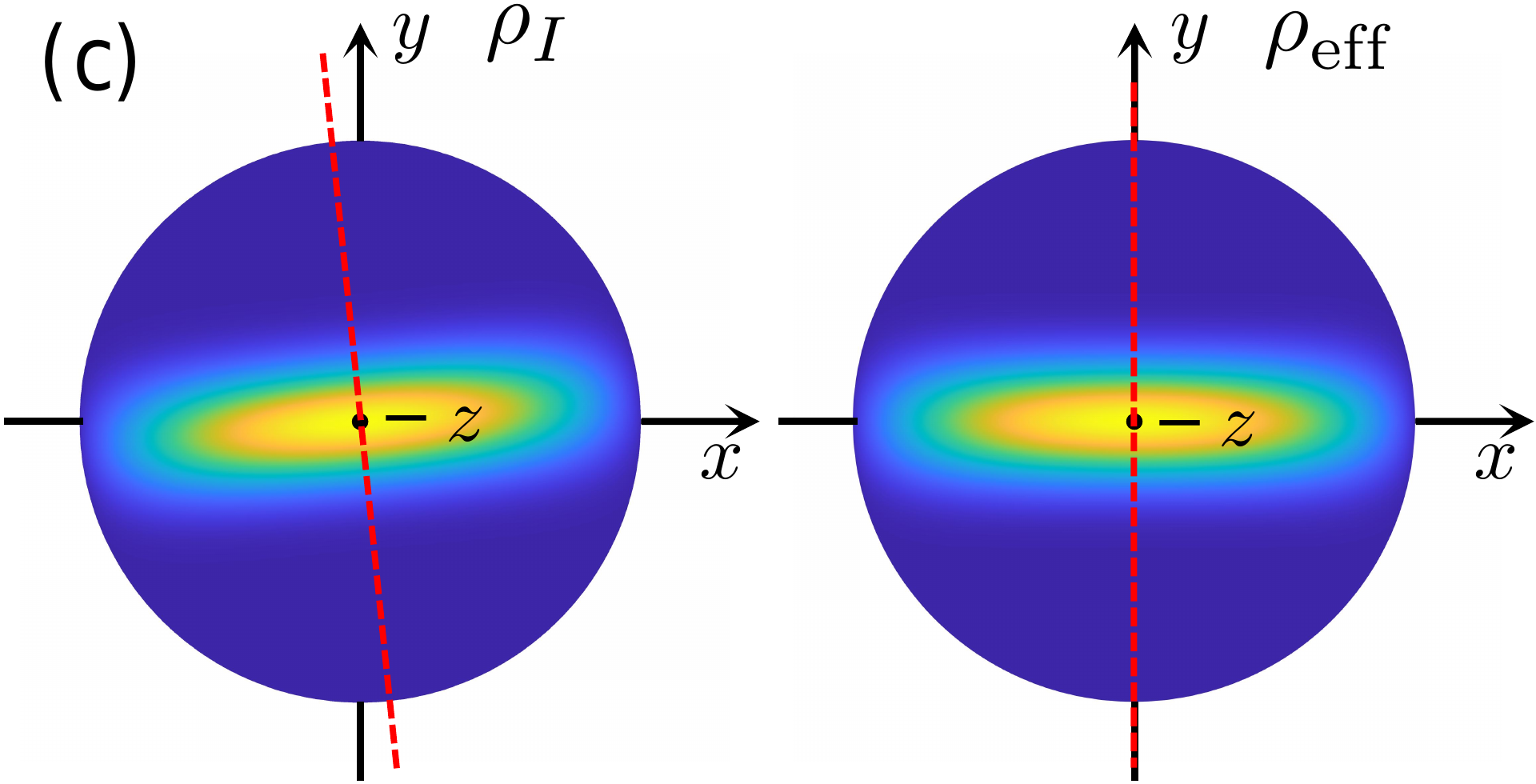}\label{fig2c}}
		\caption[Fig1]{Time evolution of (a) the parameters  $F_{I(\text{eff})}$, $\vert\langle\hat{a}_p\rangle\vert$, and (b) the spin squeezing parameter $\xi^2_R$, given by the full master equation in Eq.~(\ref{sec2eq22}) (solid curve) and the effective master equation in Eq.~(\ref{sec2eq23}) (dashed curve) for ${\kappa_{p}=\Omega=g_c}$. All other parameters are ${N=50}$, ${\delta_{s}=15g_c}$, $J=\sqrt{2}g_c$, and ${\kappa_{s}=\gamma_{c}=\gamma_{s}=0}$. (c) Husimi-$Q$ function and spin squeezing direction (red dashed curve) for the strongest spin squeezing (corresponding to the dots in (b)). Here, $\rho_I$ and $\rho_{\text{eff}}$ represent the states given by the full master equation in Eq.~(\ref{sec2eq22}) and the effective master equation in Eq.~(\ref{sec2eq23}), respectively.}
		\label{fig2}
	\end{figure}
	
	First, we assume that the dynamics of the system can be well described by the master equation in Eq.~(\ref{sec2eq3}). From the discussion below Eq.~(\ref{sec2eq17}), one can find that the strength of the generated spin squeezing is determined by the state of the pump cavity during the evolution. Therefore, the evolution of the pump cavity is worth studying. According to Eqs.~(\ref{sec2eq16}) and (\ref{sec2eq17}), the pump cavity seems to stay in a coherent state effectively during the evolution when one sets ${\{\kappa_p,\Omega\}\gg g_{\text{eff}}}$. Thus, we introduce a coherent state $\vert\beta_p\rangle_p$ whose amplitude is $\beta_p\equiv\langle \hat{a}_p\rangle_p$ (time-dependent), as a reference state, and define the fidelity $F_{I(\text{eff})}={_p\langle}\beta_p\vert\rho_{I(\text{eff})}\vert\beta_p\rangle_p$ to study the state of the pump cavity. From Fig.~\ref{fig2a}, we can obverse that both $F_{I}$ and $F_{\text{eff}}$ are always greater than $0.995$ within the evolution time of $t=300/g_c$. In other words, the pump cavity stays in the coherent state effectively during the evolution. Therefore, according to Eq.~(\ref{sec2eq17a}), the dynamics of the ensemble can be well described by the TAT model if the pump cavity is initially in a coherent state. Therefore, the protocol could give rise to a strong squeezing strength. According to Fig.~\ref{fig2b}, we find that under the chosen parameters, the protocol is able to generate a spin squeezing of $\xi^2_R\sim-9.24\,\text{dB}$ in an ensemble of $N=50$ atoms. 
	
	\begin{figure}
		\subfigure{\includegraphics[scale=0.56]{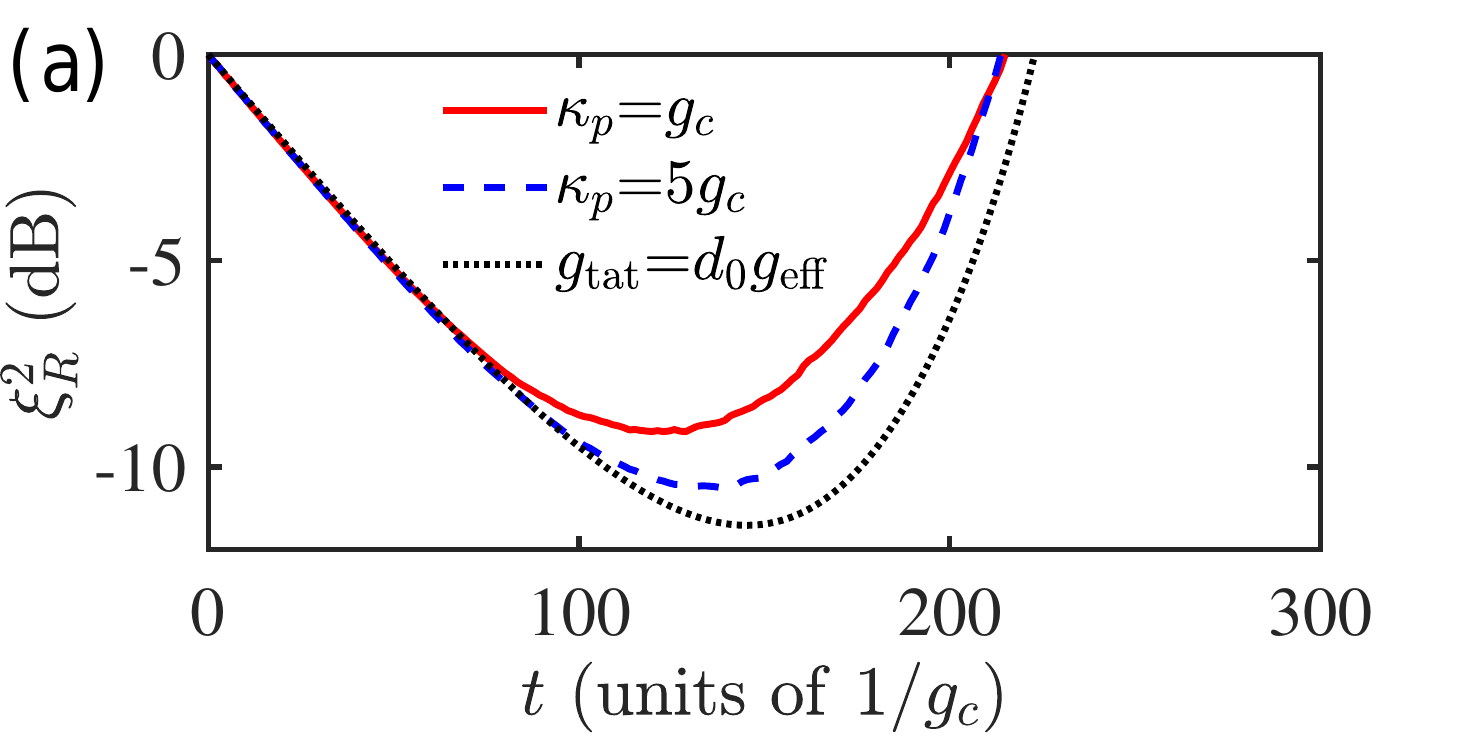}\label{fig3a}}
		\subfigure{\includegraphics[scale=0.54]{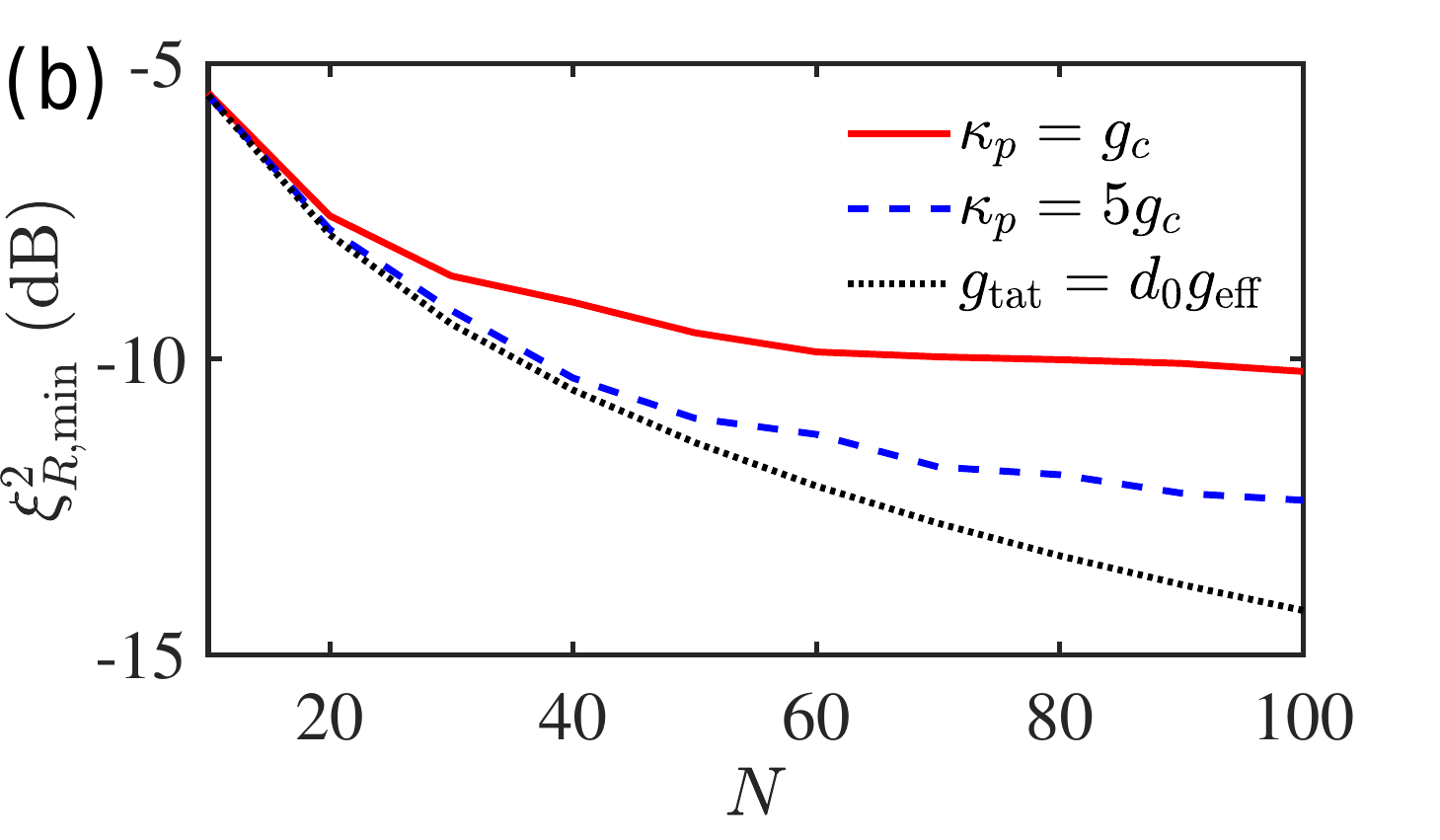}\label{fig3b}}
		\caption[Fig1]{(a) Time evolution of the spin squeezing parameter $\xi^2_R$ for an ensemble of ${N=50}$ atoms, given by the full master equation in Eq.~(\ref{sec2eq22}). (b) Minimum values of the spin squeezing parameter $\xi^2_{R,{\text{min}}}$ for different $N$, given by the effective master equation in Eq.~(\ref{sec2eq23}). We assumed that ${\kappa_{p}=g_c}$ (red solid curve) and ${\kappa_{p}=5g_c}$ (blue dashed curve) for the present protocol, and that $g_{\text{tat}}= d_0g_{\text{eff}}=ig_{\text{eff}}$ (black dotted curve) for the TAT protocol. Other parameters are ${\Omega=\kappa_{p}}$, ${\delta_{s}=15g_c}$, $J=\sqrt{2}g_c$, and $\kappa_{s}=\gamma_{c}=\gamma_{s}=0$.} 
		\label{fig3}
	\end{figure}
	
	Moreover, the squeezing direction is also an essential property of spin squeezing. From the cavity-induced TAT-like interaction and the initial coherent state of the pump cavity, ${\langle\hat{S}_x\rangle=\langle\hat{S}_y\rangle=0}$ is maintained during the evolution of the system. Therefore, the mean-spin direction is along the $z$ direction and the spin squeezing only occurs in the ${x{-}y}$ plane. To find the spin squeezing direction $\boldsymbol{n}_\perp$, we randomly choose a direction of the $x{-}y$ plane ${\boldsymbol{n}_r=\cos\theta\,\boldsymbol{e}_x+\sin\theta\,\boldsymbol{e}_y}$, where $\theta$ is the polar angle. As discussed above, the argument $\varphi$ is maintained effectively during the evolution. According to Eq.~(\ref{sec2eq17a}) and Ref.~\cite{Ma2011}, the squeezing direction $\boldsymbol{n}_\perp$ satisfies $\theta=\pi/4+\varphi/2$. Here, we demonstrate the spin squeezing direction at the moment where the strongest squeezing occurs (the dots in Fig.~\ref{fig2b}). To show the spin squeezing direction intuitively, we introduce the Husimi-$Q$ function which represents the quasiprobability distribution of any spin states \cite{Ma2011}. The Husimi-$Q$ function is defined as 
	\begin{eqnarray}\nonumber
		Q_{I(\text{eff})}=\langle \text{CSS}\vert \hat{R}^\dagger(\theta_Q,\phi_Q)\rho_{I(\text{eff})}\hat{R}(\theta_Q,\phi_Q)\vert \text{CSS}\rangle,
	\end{eqnarray}
	where $\vert\text{CSS}\rangle$ represents a CSS with all the atoms in their excited states and $\hat{R}(\theta_Q,\phi_Q)=\exp[i\theta_Q(\hat{S}_x\sin\phi_Q-\hat{S}_y\cos\phi_Q)]$ is a rotation operator \cite{Ma2011}. From Fig.~\ref{fig2c}, we find that the spin squeezing direction under the full dynamics of the system deviates slightly from the effective prediction (i.e., $y$ axis). In other words, the analysis about the spin squeezing direction is effective. Moreover, from Figs.~\ref{fig2}(a-c), we have demonstrated the validity of the effective master equation to describe the dynamics of the full system. These results show that the assumptions and the approximations we have made in Sec.~\ref{SEC2} are suitable. 
	
	\begin{figure}
		\includegraphics[scale=0.55]{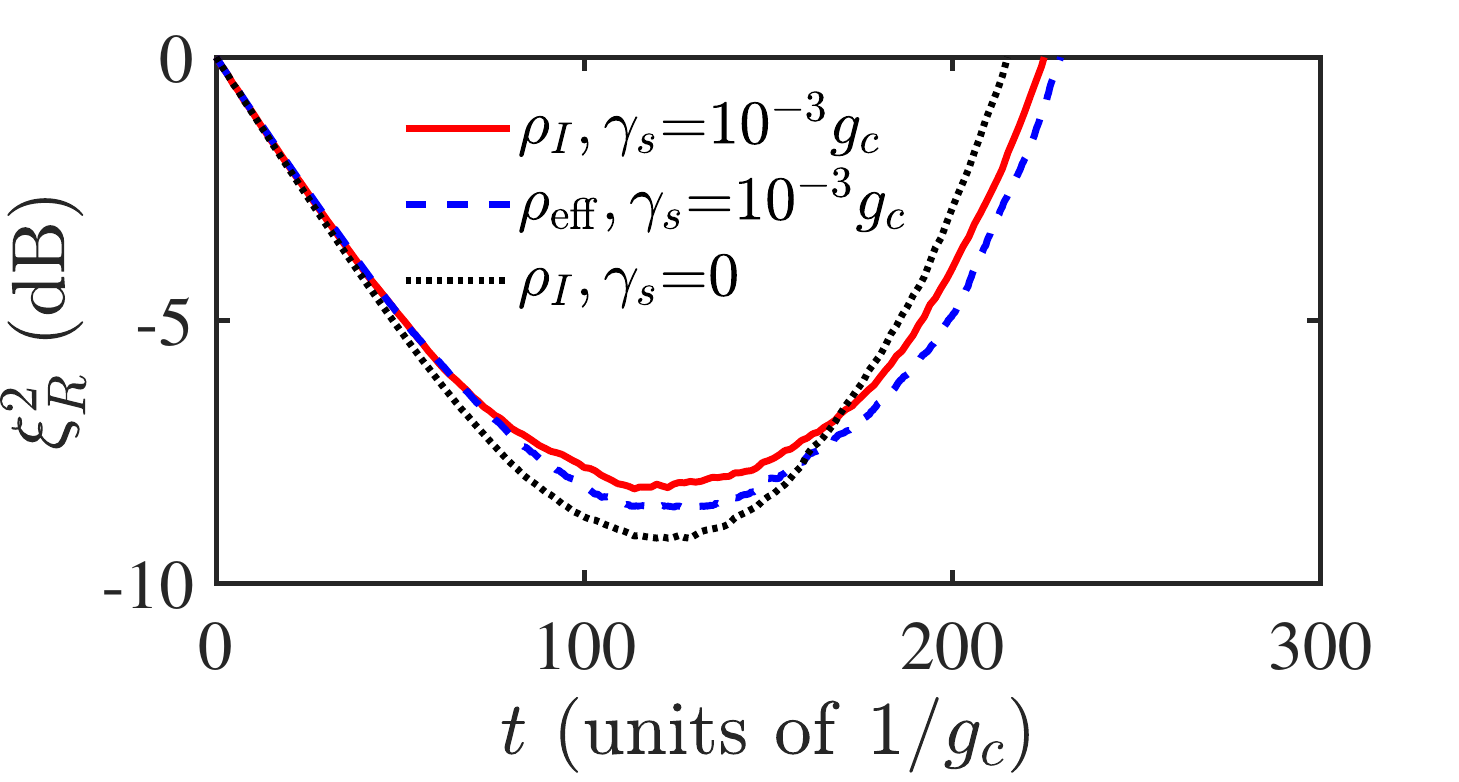}
		\caption[Fig1]{Time evolution of the spin squeezing parameter $\xi^2_R$ for different $\gamma_{s}$. The involved spontaneous emission rates $\gamma_{s}$ are set to be: ${10^{-3}g_c}$ (for $\rho_I$, red solid curve, and for $\rho_{\text{eff}}$, blue dashed curve) and $0$ (for $\rho_I$, black dotted curve). All other parameters are same as those in Fig.~\ref{fig2}.} 
		\label{fig4}
	\end{figure}
	
	\begin{figure}
		\includegraphics[scale=0.55]{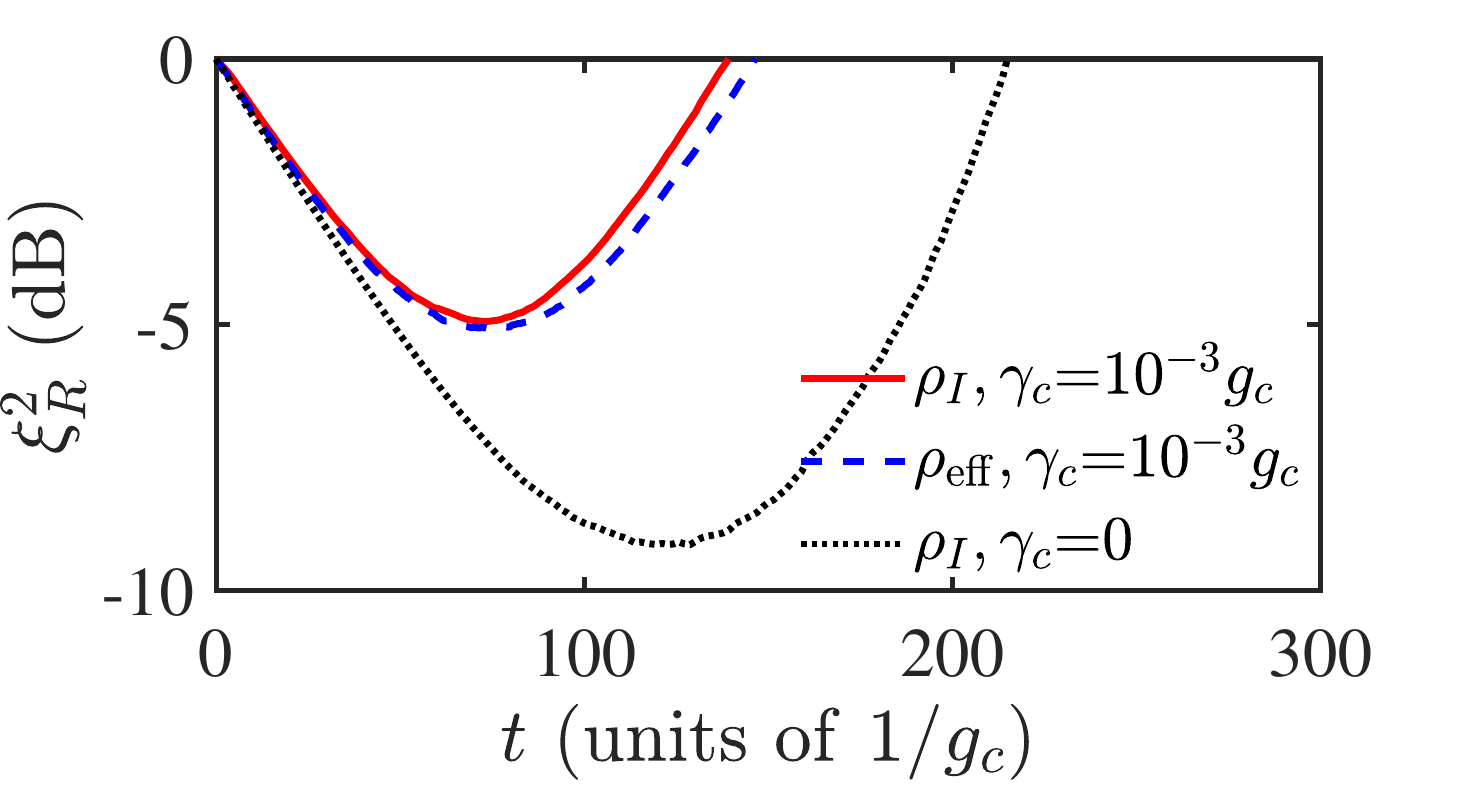}
		\caption[Fig1]{Time evolution of the spin squeezing parameter $\xi^2_R$ for different $\gamma_{c}$. The involved dephasing rates $\gamma_{c}$ are set to be: ${10^{-3}g_c}$ (for $\rho_I$, red solid curve, and for $\rho_{\text{eff}}$, blue dashed curve) and ${0}$ (for $\rho_I$, black dotted curve). All other parameters are same as those in Fig.~\ref{fig2}.} 
		\label{fig5}
	\end{figure}
	
	According to the mechanism of the cavity-induced TAT-like interaction in Sec.~\ref{SEC2},  when the condition $\{\kappa_{p},\Omega\}\gg g_\text{eff}$ is satisfied, the dynamics of the system is consistent with that of the TAT model. To confirm this, we now introduce the TAT protocol, as a reference. The Hamiltonian in the TAT protocol is given by
	\begin{eqnarray}\label{sec3eq2}
		\hat{H}_{\text{TAT}}=g_{\text{tat}}\hat{S}^2_++g^*_{\text{tat}}\hat{S}^2_-,
	\end{eqnarray}
	where ${g_{\text{tat}}=d_0g_{\text{eff}}}$. Compared to the present protocol, the TAT protocol is equivalent to a case where the pump cavity stays in the quasi-steady state during the evolution. Thereafter, we numerically compare the present protocol and the TAT protocol. Figure \ref{fig3a} plots the evolution of $\xi_R^2$ verse $t$, showing that a larger $\kappa_p$ increases the similarity between the present protocol and the TAT protocol. In particular, all the curves are in clear agreement with each other at the beginning of the evolution. The reason is that, from Fig.~\ref{fig2a}, we can obtain that the pump cavity is in the quasi-steady state effectively at the beginning of the evolution. From Fig.~\ref{fig3b}, one can find that for a small ensemble, the present protocol can generate the same spin squeezing strength as the TAT protocol, e.g., ${\Omega=\kappa_{p}=5g_c}$ for ${N=10}$. Moreover, it is also seen that, as the number of atoms increases, the minimum value of the spin squeezing parameter $\xi^2_{R,\min}$ is decreased for both the present protocol and the TAT protocol. However, an increasing in the number of atoms causes the strongest spin squeezing generated with the present protocol to gradually deviate from that generated with the TAT protocol. The reason is that the generation of spin squeezing in an ensemble with more atoms generally leads more atoms to be excited, which would cause the state of the pump cavity to deviate more from the quasi-steady coherent state. Moreover, according to Figs.~\ref{fig3a} and \ref{fig3b}, a larger $\kappa_p$ results in a stronger spin squeezing. This means that a setup with a larger single-photon dissipation of the pump cavity and a stronger external driving field is more efficient for the present protocol to generate a stronger spin squeezing. 
	
	Then, we study the influence of the spontaneous emission of the atoms and the collective dephasing of the ensemble. From Fig.~\ref{fig4}, we find that the effect of spontaneous emission prolongs the duration of the squeezing at the expense of reducing the squeezing strength slightly. Meanwhile, it can be seen from Fig.~\ref{fig5} that, at the same intensity, the influence of the collective dephasing of the ensemble on the generated spin squeezing is much greater than that caused by the atomic spontaneous emission. The essential reason is that the collective dephasing would affect the relative phases between the energy levels of the ensemble, so that the direction of the squeezing gradually becomes chaotic during the evolution.
			
	\begin{figure}
		\includegraphics[scale=0.55]{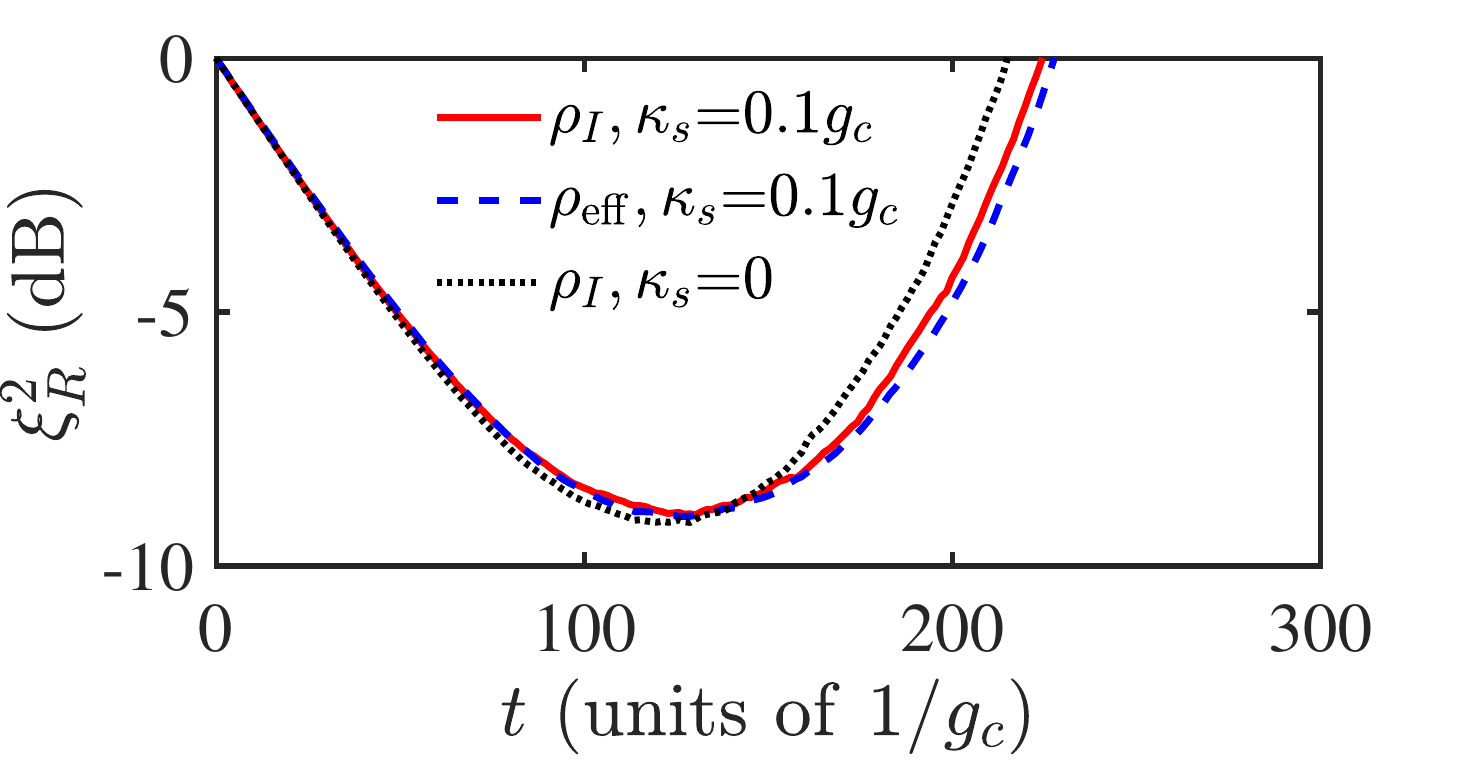}
		\caption[Fig1]{Time evolution of the spin squeezing parameter $\xi^2_R$ for different $\kappa_{s}$. The involved signal cavity decay rates $\kappa_{s}$ are set to be: ${0.1g_c}$ (for $\rho_I$, red solid curve, and for $\rho_{\text{eff}}$, blue dashed curve) and ${0}$ (for $\rho_I$, black dotted curve). All other parameters are same as those in Fig.~\ref{fig2}.} 
		\label{fig6}
	\end{figure}	
		 
	Though the signal cavity is decoupled from the effective dynamics described by Eq.~(\ref{sec2eq23}), the signal cavity decay affects inevitably the full dynamics of the system in reality. Here, we discuss the influence of the signal cavity decay on the system. With the help of the theory of the adiabatic elimination \cite{PhysRevA.85.032111,PhysRevA.97.032328}, we can replace $\gamma_{s}\mathcal{L}(\hat{a}_s)\rho$ with  $J^2p_\gamma\mathcal{L}({\hat{a}_p})\rho+g^2p_\gamma\mathcal{L}(\hat{S}_-)\rho$, where ${p_\gamma=4\gamma_{s}/(4\delta^2_{s}+\gamma^2_{s})}$, in the master equation in Eqs.~(\ref{sec2eq22}) and (\ref{sec2eq23}) (see more details in \ref{APPa}). Therefore, the effect of the signal cavity decay causes an additional single-photon dissipation of the pump cavity and an additional collective spontaneous emission of the atoms. From Sec.~\ref{SEC2}, the condition ${\kappa_{p}\gg\ J^2p_\gamma}$ is well satisfied, which means that the adiabatic effect on the pump cavity may be not obvious. Then, the signal cavity decay mainly produces an additional collective spontaneous emission. This analysis can be verified indirectly by comparing the evolution of the spin squeezing parameter $\xi^2_R$ in Figs.~\ref{fig4} and \ref{fig6}.
	
	\subsection{WITHOUT DRIVING THE PUMP CAVITY}\label{SEC3b}
	
	\begin{figure}
		\subfigure{\includegraphics[scale=0.54]{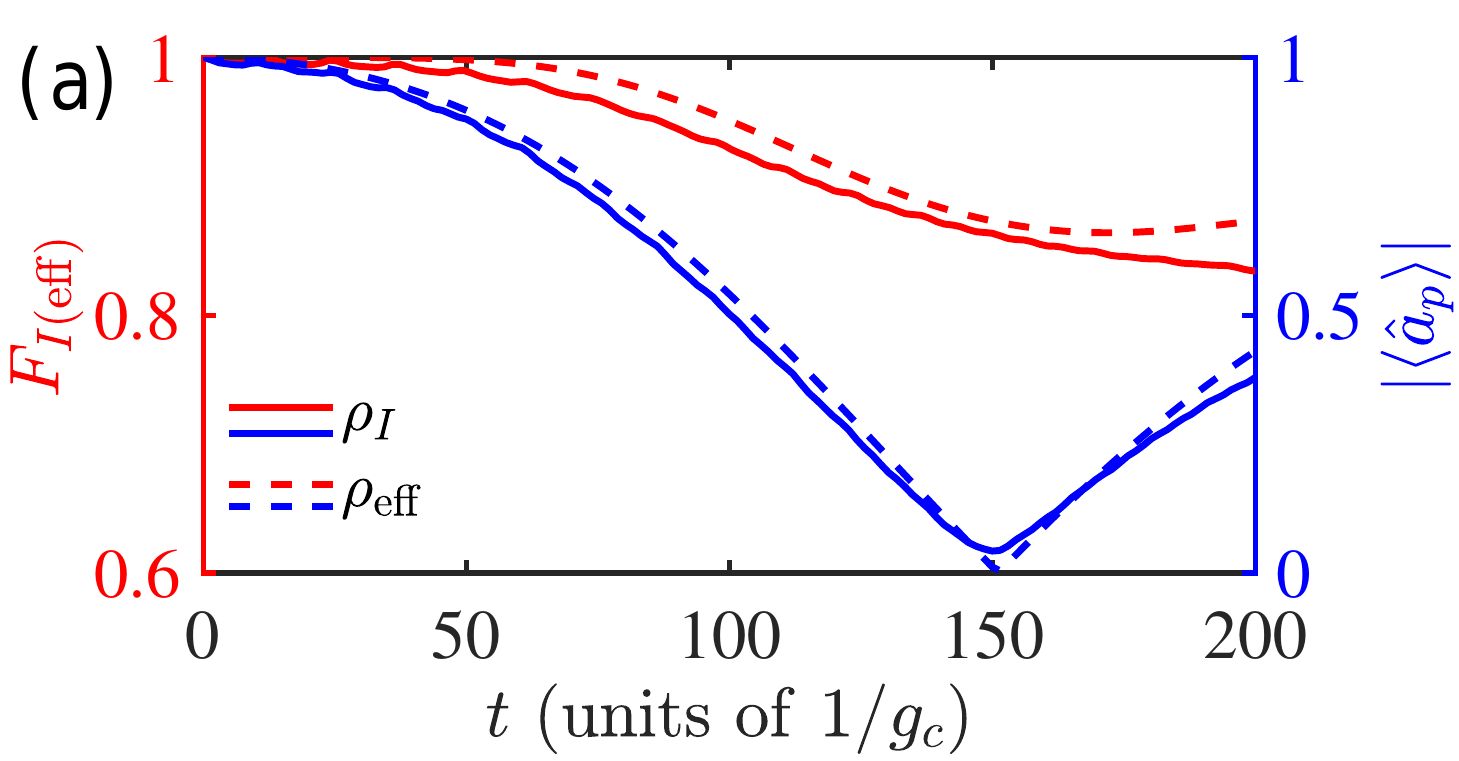}\label{fig7a}}
		\subfigure{\includegraphics[scale=0.65]{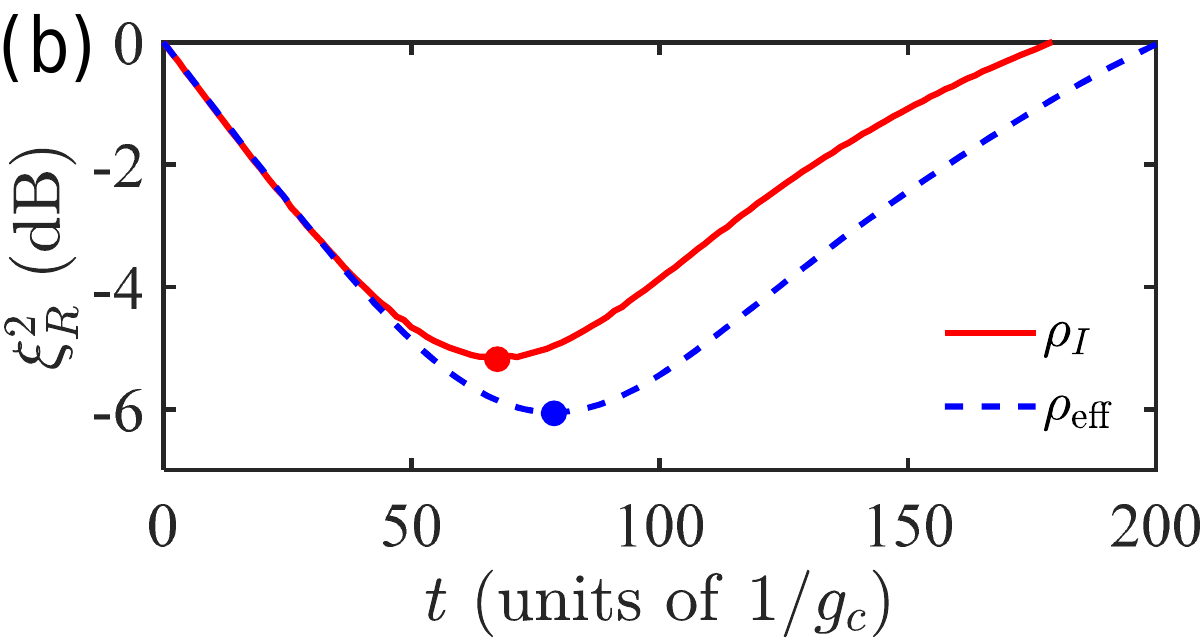}\label{fig7b}}
		\subfigure{\includegraphics[scale=0.40]{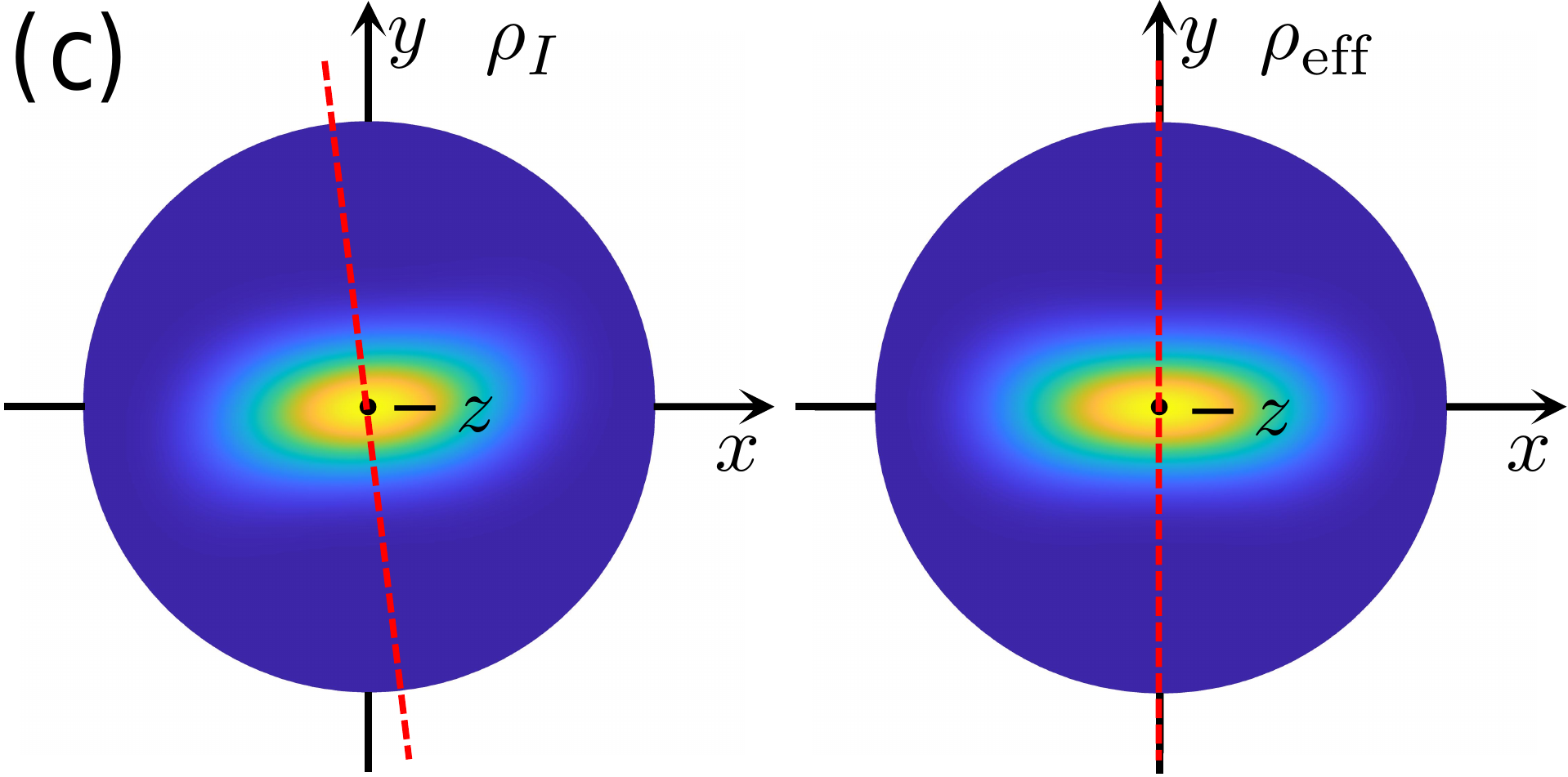}\label{fig7c}}
		\caption[Fig1]{Time evolution of (a) the parameters  $F_{I(\text{eff})}$, $\vert\langle\hat{a}_p\rangle\vert$, and (b) the spin squeezing parameter $\xi^2_R$, given by the full master equation in Eq.~(\ref{sec2eq22}) (the solid curve) and the effective master equation in Eq.~(\ref{sec2eq23}) (the dashed curve) for ${\alpha=i}$. All other parameters are ${N=50}$, ${\delta_{s}=15g_c}$, $J=\sqrt{2}g_c$, and $\Omega=\kappa_{p}=\kappa_{s}=\gamma_{c}=\gamma_{s}=0$. (c) Husimi-$Q$ function and spin squeezing direction (the red dashed curve) for the strongest spin squeezing (corresponding to the dots in (b)).} 
		\label{fig7}
	\end{figure}
	
	In this subsection, we discuss the property of the generated spin squeezing without driving the pump cavity. Here, the effective Hamiltonian in {Eq.~(\ref{sec2eq16})} is given by
	\begin{equation}\label{sec3eq3}
		\hat{H}'_\text{eff}=g'_\text{eff}(\hat{a}_p\hat{S}^2_++\hat{a}^\dagger_p\hat{S}^2_-),
	\end{equation}
	where $g'_\text{eff}=g^2J/{\delta_s^2}$. 
		
	\begin{figure}
		\includegraphics[scale=0.55]{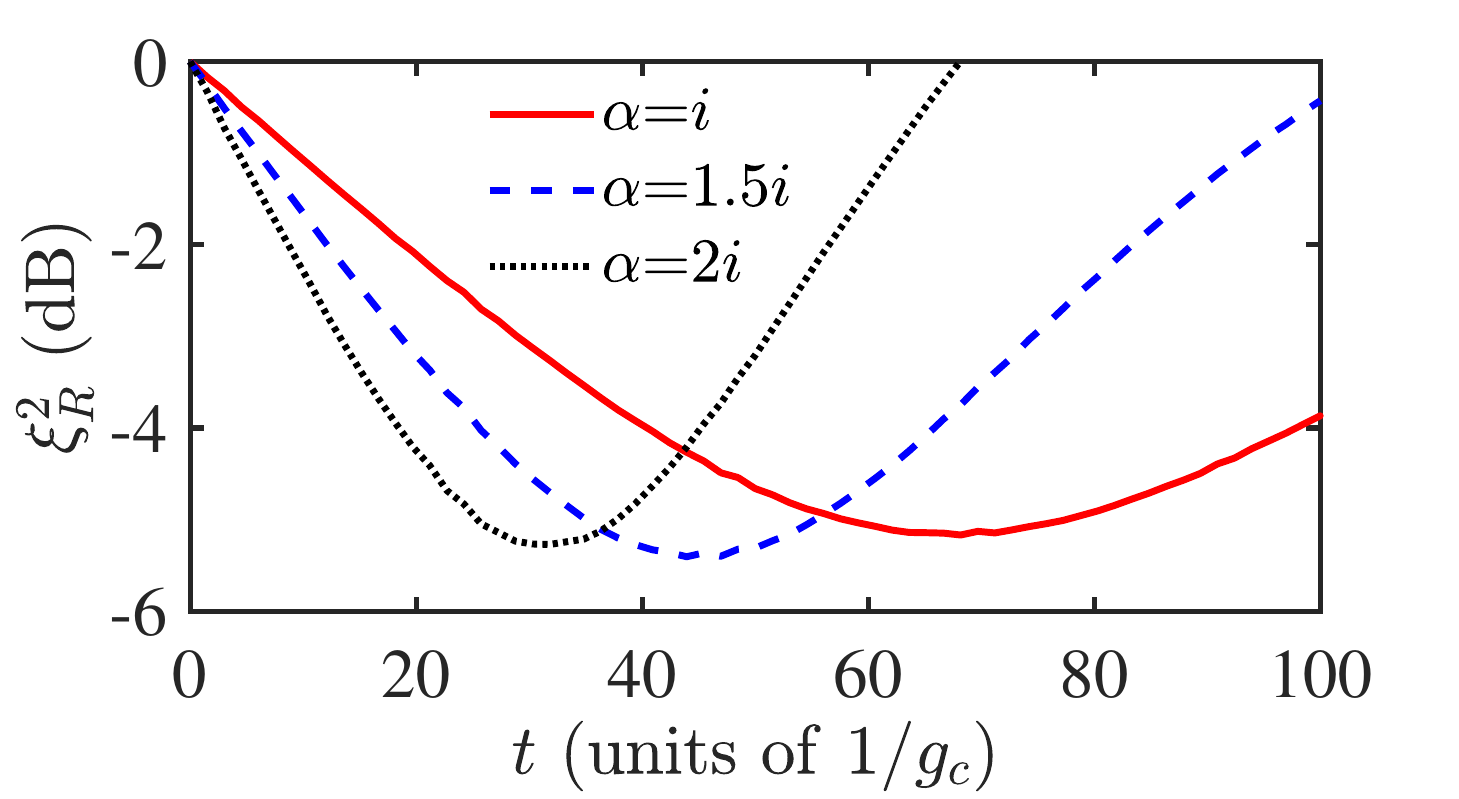}
		\caption[Fig1]{Time evolution of the spin squeezing parameter $\xi^2_R$ for different $\alpha$, given by the full master equation in Eq.~(\ref{sec2eq22}). The involved amplitudes of the initial coherent state $\alpha$ are set to be: ${i}$ (red solid curve), ${1.5i}$ (blue dashed curve), and ${2i}$ (black dotted curve). All other parameters are same as those in Fig.~\ref{fig7}.} 
		\label{fig8a}
	\end{figure}	
		
	First, we study the generation of spin squeezing in the ideal case of no decoherence involved. Similar to the Sec.~\ref{SEC3a}, we start from studying the coherence properties of the pump cavity with a coherent state as a reference state. Without the driving, the amplitude of the reference state is decreased sharply in the early stage, as shown in Fig.~\ref{fig7a}. This indicates that the initial coherent state of the pump cavity is destroyed, which, according to Eq.~(\ref{sec2eq17a}), limits the strength of the generated spin squeezing. As shown in Fig.~\ref{fig7b}, the protocol is able to generate a spin squeezing of ${\xi^2_R\sim-5.19\,\text{dB}}$ in an ensemble of ${N=50}$ atoms. Meanwhile, from Figs.~\ref{fig7a} and \ref{fig7b}, we find that the fidelity of the reference coherent state exceeds $0.98$ before the strongest squeezing occurs. In other words, the pump cavity is in a coherent state approximately. Therefore, the prediction of the spin squeezing direction along $\theta=\pi/4+\varphi/2$ in the $x{-y}$ plane is still valid. From Fig.~\ref{fig7c}, the spin squeezing direction based on the full master equation only has a small deviation from the effective prediction (i.e., along the $y$ axis). Moreover, from Figs.~\ref{fig7a} and \ref{fig7c}, the effective master equation is also still valid to describe the system dynamics. However, the effective master equation, when used to predict the evolution of $\xi^2_R$, has a big deviation due to the sensitivity of $\xi^2_R$ to the parameters. Therefore, the following numerical simulations in this subsection are all based on the full master equation.
	
	According to Figs.~\ref{fig7a} and \ref{fig7b}, an increase in the amplitude of the initial coherent state of the pump cavity might lead to a faster and stronger spin squeezing. Here, we numerically study this deduction and plot the corresponding results in Fig.~\ref{fig8a}. From Fig.~\ref{fig8a}, a larger-amplitude coherent state prepared in the pump cavity accelerates the process of generating spin squeezing. The reason is that according to Eq.~(\ref{sec2eq17a}), an increase in the amplitude of the coherent state in the pump cavity increases the coupling strength of the nonlinear interaction between the atoms. But at the same time, a larger-amplitude coherent state in the pump cavity also involves a violation of the assumption $\delta_{s}\gg 2|\alpha|J$. This results in a decrease of the generated squeezing strength, as shown in Fig.~\ref{fig8a}.
	
	\begin{figure}
		\includegraphics[scale=0.55]{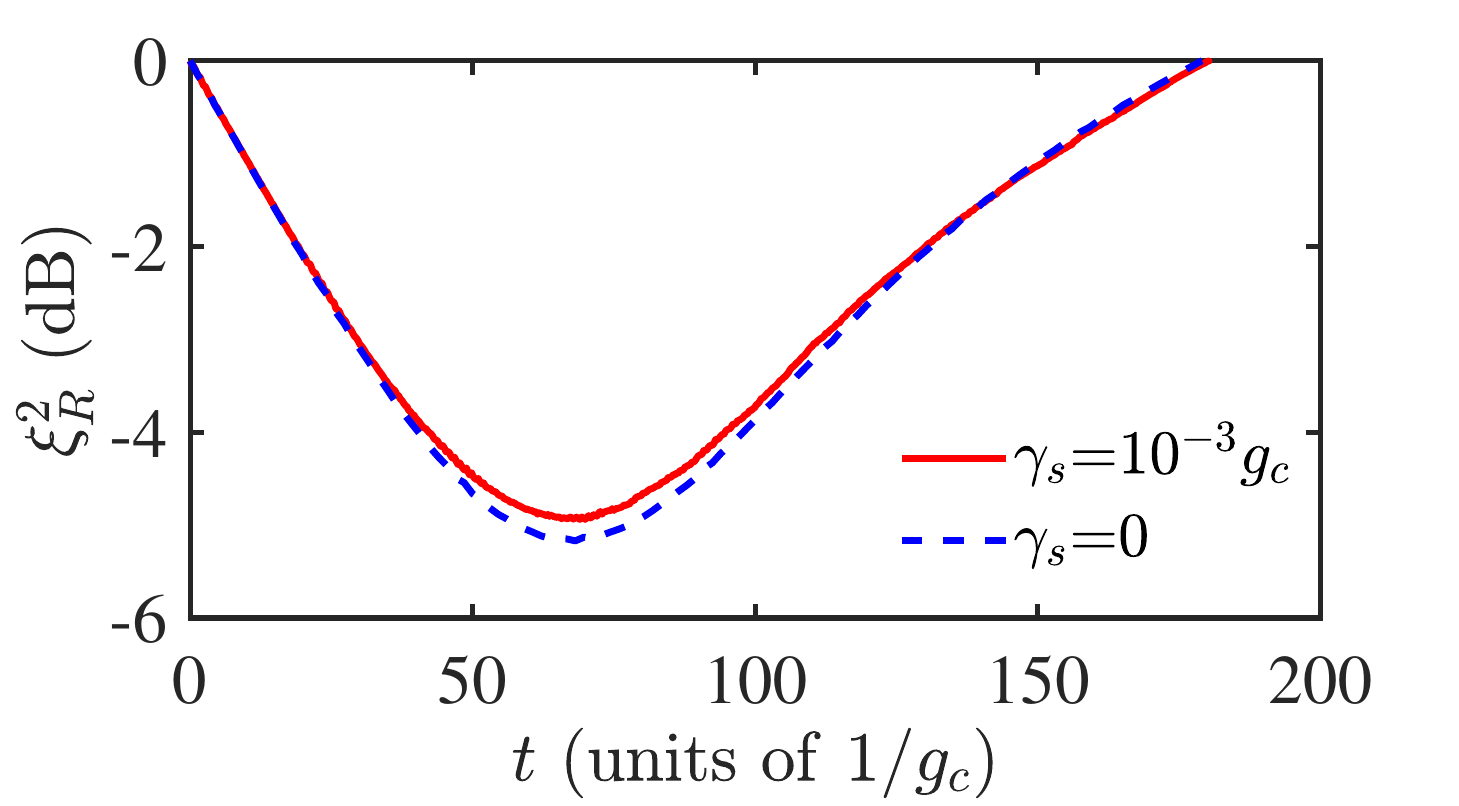}
		\caption[Fig1]{Time evolution of the spin squeezing parameter $\xi^2_R$ for different $\gamma_{s}$, given by the full master equation in Eq.~(\ref{sec2eq22}). The involved atomic spontaneous emission rates $\gamma_{s}$ are set to be: ${10^{-3}g_c}$ (red solid curve) and ${0}$ (blue dashed curve). All other parameters are same as those in Fig.~\ref{fig7}.} 
		\label{fig8}
	\end{figure}
	
	\begin{figure}
		\includegraphics[scale=0.55]{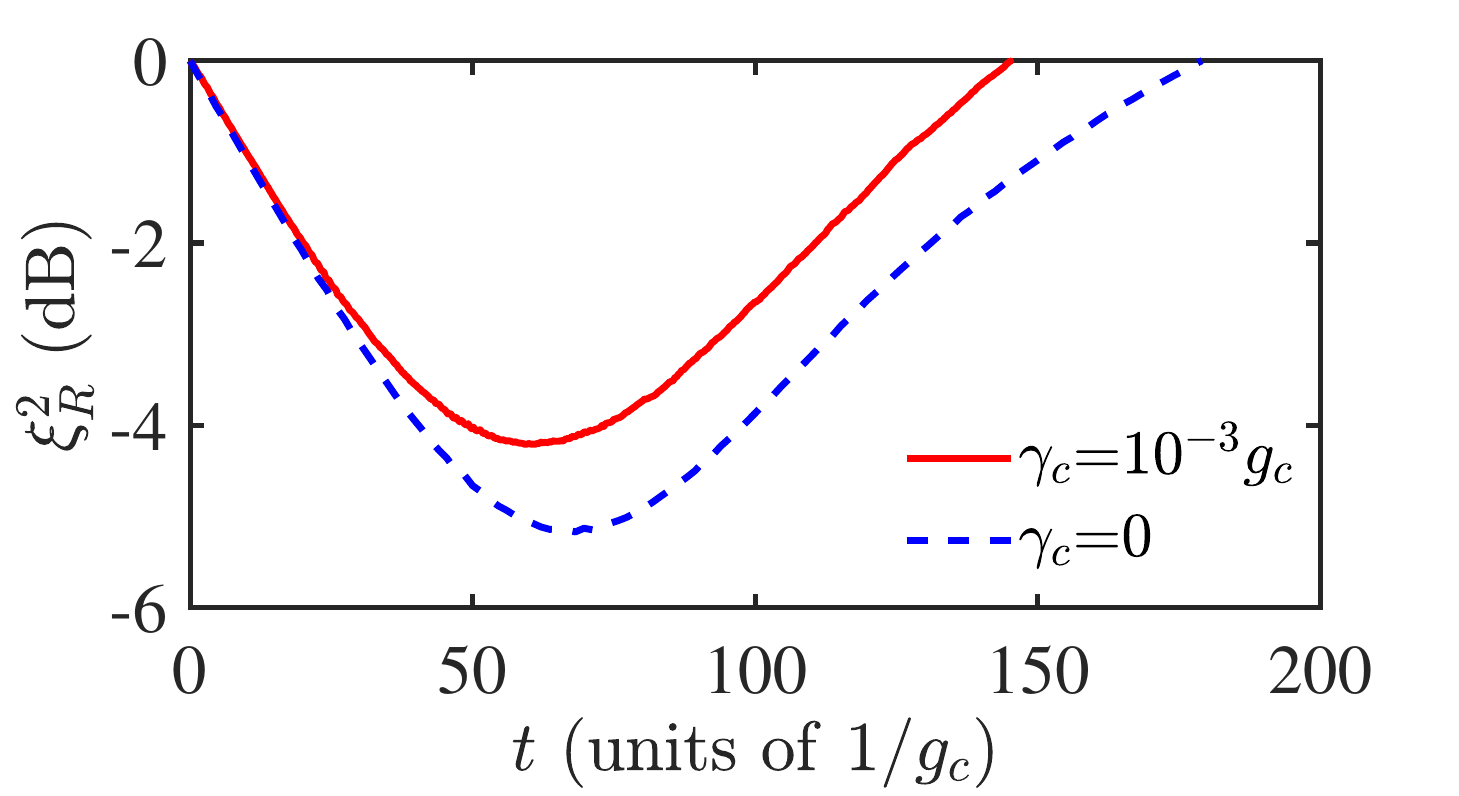}
		\caption[Fig1]{Time evolution of the spin squeezing parameter $\xi^2_R$ for different $\gamma_{c}$, given by the full master equation in Eq.~(\ref{sec2eq22}). The involved collective dephasing rates $\gamma_{c}$ are set to be: ${10^{-3}g_c}$ (red solid curve) and ${0}$ (blue dashed curve). All other parameters are same as those in Fig.~\ref{fig7}.} 
		\label{fig9}
	\end{figure}
	
	It is worth studying further the effects of the spontaneous emission of the atoms and the collective dephasing of the ensemble on the system. As shown in Fig.~\ref{fig8}, the spontaneous emission of the atoms slightly influence on the evolution of spin squeezing. Compared to the case with driving the pump cavity in Sec.~\ref{SEC3a}, the atomic spontaneous emission cannot prolong the duration of spin squeezing in the case of no pump cavity driving. According to Figs.~\ref{fig8} and \ref{fig9}, when setting $\gamma_{s}=\gamma_{c}$, the dephasing of the ensemble has a greater effect on the generation of spin squeezing than the atomic spontaneous emission. Meanwhile, from Figs.~\ref{fig5} and \ref{fig9}, it can be found that in the case of no pump cavity driving, the collective dephasing of the ensemble has a weaker influence on the generation of spin squeezing than that in the case with driving the pump cavity.
	
	\begin{figure}
		\includegraphics[scale=0.55]{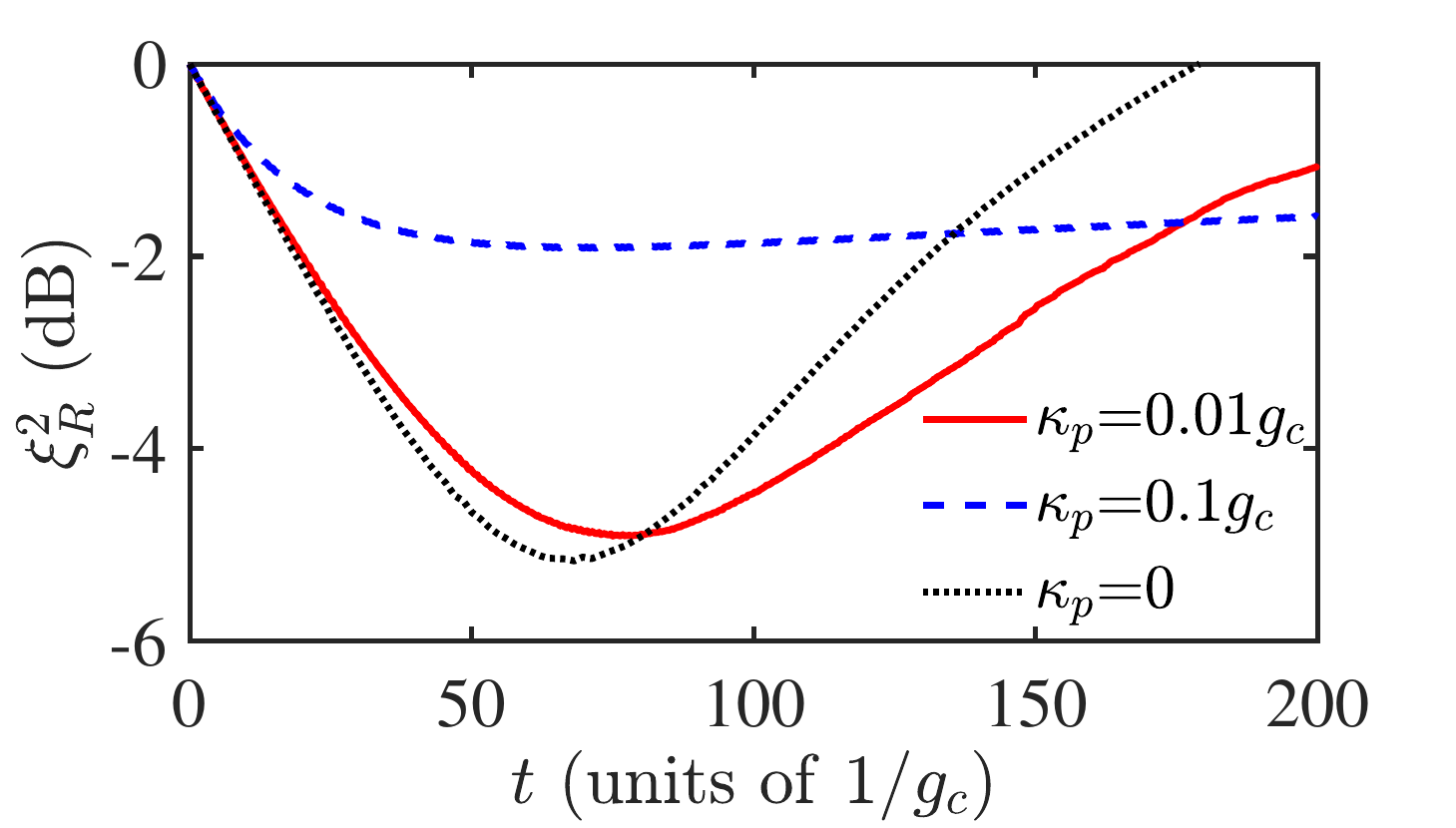}
		\caption[Fig1]{Time evolution of the spin squeezing parameter $\xi^2_R$ for different $\kappa_{p}$, given by the full master equation in Eq.~(\ref{sec2eq22}). The involved single-photon dissipation rates of the pump cavity $\kappa_{p}$ are set to be: $0.01g_c$ (red solid curve), $0.1g_c$ (blue dashed curve), and $0$ (black dotted curve). All other parameters are same as those in Fig.~\ref{fig7}.} 
		\label{fig11}
	\end{figure}

	\begin{figure}
		\includegraphics[scale=0.55]{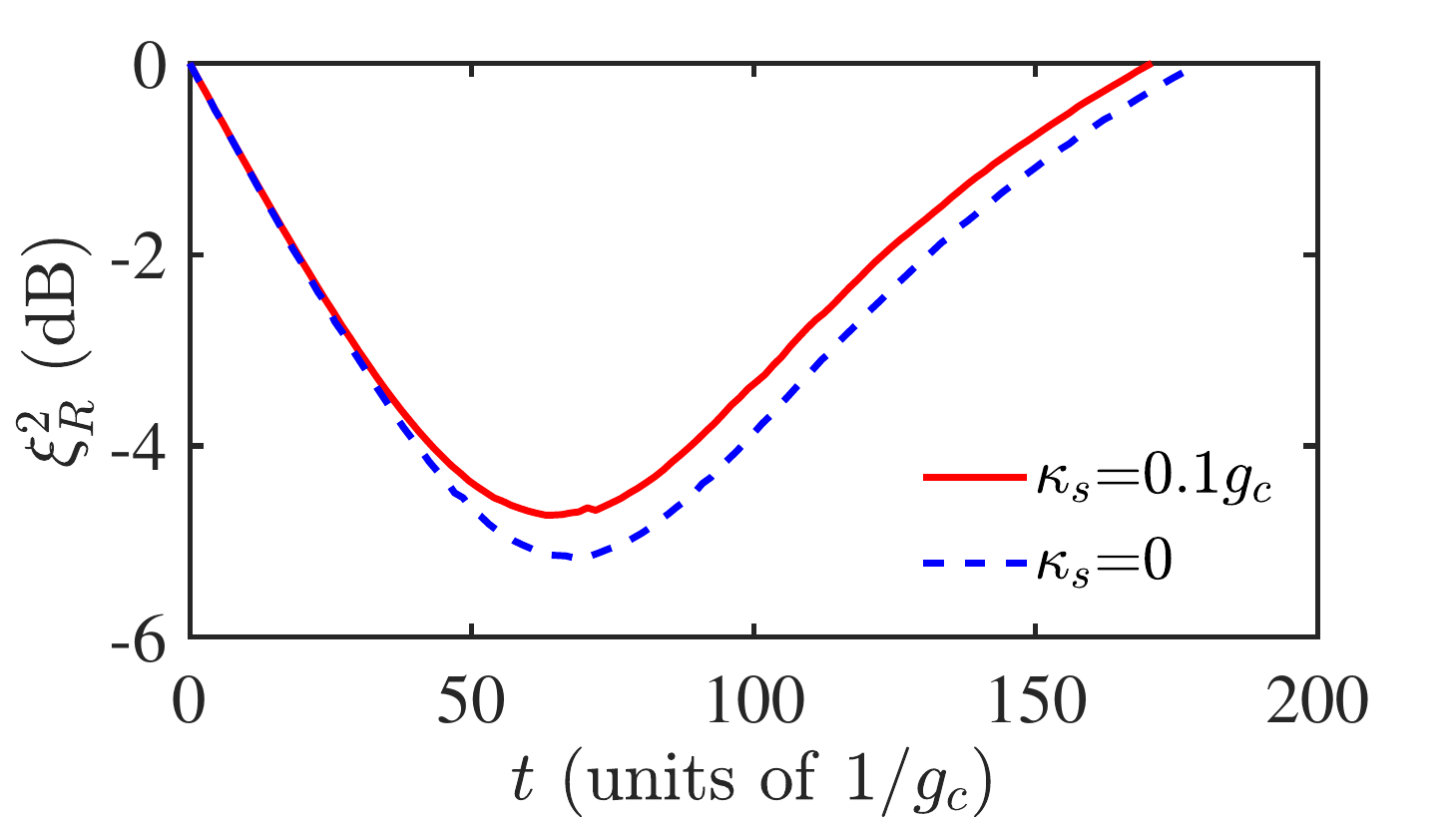}
		\caption[Fig1]{Time evolution of the spin squeezing parameter $\xi^2_R$ for different $\kappa_{s}$, given by the full master equation in Eq.~(\ref{sec2eq22}). The involved single-photon dissipation rates  of the signal cavity $\kappa_{s}$ are set to be: $0.1g_c$ (red solid curve) and $0$ (blue dashed curve). All other parameters are same as those in Fig.~\ref{fig7}.} 
		\label{fig10}
	\end{figure}
	
	Furthermore, the effect of the cavity decay on the system also needs to be investigated. Here, the pump cavity decay becomes an undesired physical process.  According to Fig.~\ref{fig11}, the pump cavity decay can significantly prolong the duration of squeezing, but at the same time, it can also clearly reduce the strength of squeezing. This indicates that the protocol in the case without driving the pump cavity is particularly suitable for a system with a high-quality pump cavity. Meanwhile, as discussed in Sec.~\ref{SEC3a}, the influence of the signal cavity decay is equivalent to introducing an extra pump cavity decay and an extra atomic collective spontaneous emission. It is worth noting that the extra pump cavity decay cannot be ignored like in the case with driving the pump cavity. Thus, the robustness of the generated spin squeezing to the signal cavity decay would be reduced, which is shown in Fig.~\ref{fig10}.
		
	\section{EXPERIMENTAL FEASIBILITY}\label{SEC4}
	
	\begin{figure}
		\includegraphics[scale=0.35]{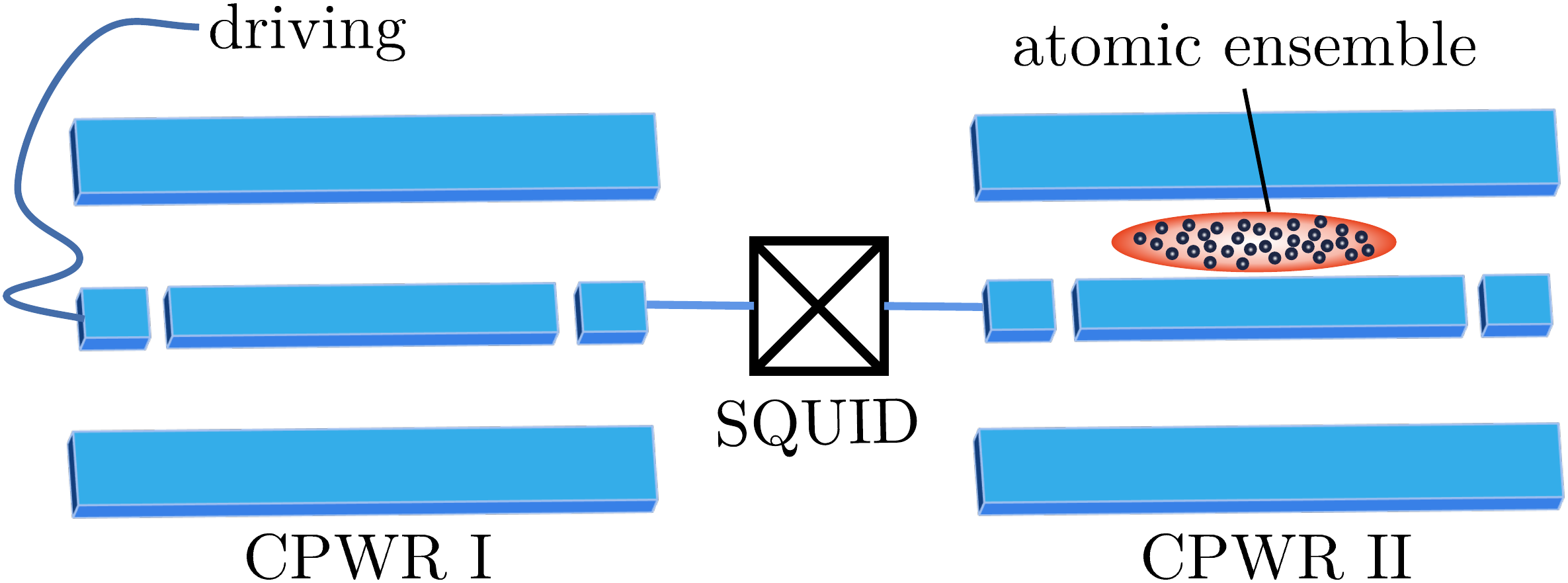}
		\caption[Fig1]{Schematic for a possible architecture for the protocol. The CPWR I and the CPWR II are coupled to the SQUID. The atomic ensemble is placed above the CPWR II and is coupled to the magnetic mode of the CPWR II. An extra driving is applied to the CPWR I.} 
		\label{fig12}
	\end{figure}
	\begin{table*}[ht]
		\centering
		\caption{Experimental feasible parameters and the minimum of spin squeezing parameters}
		\label{table1}
	\begin{threeparttable}
		\setlength\tabcolsep{4pt}
		\begin{tabular}{ccccccccccccccccccccc}  
			\hline\\  [-9pt]
			\hline
			\multicolumn{1}{c}{\ } & \multicolumn{1}{c}{\ } & \multicolumn{1}{c}{\ } & \multicolumn{1}{c}{\ } & \multicolumn{1}{c}{} & \multicolumn{1}{c}{\ } & \multicolumn{1}{c}{\ } & \multicolumn{1}{c}{} &\multicolumn{1}{c}{\ } & \multicolumn{1}{c}{\ } & \multicolumn{1}{c}{} &\multicolumn{1}{c}{\ } & \multicolumn{1}{c}{\ } & \multicolumn{1}{c}{} & \multicolumn{1}{c}{\ } &	\multicolumn{1}{c}{\ } & \multicolumn{1}{c}{} &\multicolumn{1}{c}{\ } & \multicolumn{1}{c}{\ } & \multicolumn{1}{c}{} &\multicolumn{1}{c}{\ } \\
			
			\multicolumn{1}{c}{\ } & \multicolumn{1}{c}{\ } & \multicolumn{1}{c}{\ } & \multicolumn{18}{c}{Type of ensemble}\\
			
			\cmidrule{4-21}
			
			\multicolumn{1}{c}{} & \multicolumn{1}{c}{\multirow{2}{*}{Parameter}} & \multicolumn{1}{c}{} & \multicolumn{1}{c}{} & \multicolumn{7}{c}{\multirow{2}{*}{Rb atoms}} & \multicolumn{1}{c}{} & \multicolumn{1}{c}{} & \multicolumn{7}{c}{\multirow{2}{*}{NV centers}} & \multicolumn{1}{c}{} \\
			\multicolumn{1}{c}{} & \multicolumn{2}{c}{} & \multicolumn{9}{c}{} & \multicolumn{9}{c}{}\\[-7pt]
			
			\cmidrule(r){1-3} \cmidrule(r){4-12} \cmidrule{13-21}
			
			\multicolumn{1}{c}{} & \multicolumn{2}{c}{} & \multicolumn{9}{c}{} & \multicolumn{9}{c}{}\\[-7pt]
			
			\multicolumn{1}{c}{} & \multicolumn{1}{c}{encoded states} & \multicolumn{1}{c}{} & \multicolumn{1}{c}{} & \multicolumn{7}{c}{$\vert1,\,-1\rangle\leftrightarrow\vert2,\,1\rangle$}  & \multicolumn{1}{c}{} & \multicolumn{1}{c}{} & \multicolumn{7}{c}{$\vert m_s=0\rangle\leftrightarrow\vert m_s=-1\rangle$} & \multicolumn{1}{c}{}\\[4pt]
			
			\multicolumn{1}{c}{} & \multicolumn{1}{c}{number of spins} & \multicolumn{1}{c}{} & \multicolumn{1}{c}{} & \multicolumn{7}{c}{\multirow{1}{*}{$10^6$}} & \multicolumn{1}{c}{} &\multicolumn{1}{c}{} & \multicolumn{7}{c}{\multirow{1}{*}{$10^{12}$}} & \multicolumn{1}{c}{}\\ [4pt]
			
			\multicolumn{1}{c}{} & \multicolumn{1}{c}{spin frequency} & \multicolumn{1}{c}{} & \multicolumn{1}{c}{} & \multicolumn{7}{c}{\multirow{2}{*}{6.8324}} & \multicolumn{1}{c}{} & \multicolumn{1}{c}{} & \multicolumn{7}{c}{\multirow{2}{*}{2.6899}} & \multicolumn{1}{c}{}\\
			\multicolumn{1}{c}{} & \multicolumn{1}{c}{$\omega_{q}/2\pi$\,(GHz)} & \multicolumn{1}{c}{} & \multicolumn{1}{c}{} & \multicolumn{7}{c}{} & \multicolumn{1}{c}{} & \multicolumn{1}{c}{} & \multicolumn{7}{c}{} & \multicolumn{1}{c}{}\\ [4pt]
			
			\multicolumn{1}{c}{} & \multicolumn{1}{c}{spin spontaneous emission rate} & \multicolumn{1}{c}{} & \multicolumn{1}{c}{} & \multicolumn{7}{c}{\multirow{2}{*}{---}} & \multicolumn{1}{c}{} & \multicolumn{1}{c}{} & \multicolumn{7}{c}{\multirow{2}{*}{---}} &  \multicolumn{1}{c}{}\\
			\multicolumn{1}{c}{} & \multicolumn{1}{c}{$\gamma_{s}/2\pi$\,(kHz)} & \multicolumn{1}{c}{} & \multicolumn{1}{c}{} & \multicolumn{7}{c}{} & \multicolumn{1}{c}{}  & \multicolumn{1}{c}{} & \multicolumn{7}{c}{} &  \multicolumn{1}{c}{}\\ [4pt]
			
			\multicolumn{1}{c}{} & \multicolumn{1}{c}{spin collective dephasing rate} & \multicolumn{1}{c}{} & \multicolumn{1}{c}{} & \multicolumn{7}{c}{\multirow{2}{*}{---}} & \multicolumn{1}{c}{} & \multicolumn{1}{c}{} & \multicolumn{7}{c}{\multirow{2}{*}{0.26}} & \multicolumn{1}{c}{}\\
			\multicolumn{1}{c}{} & \multicolumn{1}{c}{$\gamma_{c}/2\pi$\,(kHz)} & \multicolumn{1}{c}{} & \multicolumn{1}{c}{} & \multicolumn{7}{c}{} & \multicolumn{1}{c}{} & \multicolumn{1}{c}{} & \multicolumn{7}{c}{} &  \multicolumn{1}{c}{}\\ [4pt]
			
			\multicolumn{1}{c}{} & \multicolumn{1}{c}{collective coupling strength} & \multicolumn{1}{c}{} & \multicolumn{1}{c}{} & \multicolumn{7}{c}{\multirow{2}{*}{40}} & \multicolumn{1}{c}{} & \multicolumn{1}{c}{} & \multicolumn{7}{c}{\multirow{2}{*}{1.200$\times10^4$}} & \multicolumn{1}{c}{}\\
			\multicolumn{1}{c}{} & \multicolumn{1}{c}{$g_c/2\pi$\,(kHz)} & \multicolumn{1}{c}{} & \multicolumn{1}{c}{} & \multicolumn{7}{c}{} & \multicolumn{1}{c}{} & \multicolumn{1}{c}{} & \multicolumn{7}{c}{} &  \multicolumn{1}{c}{}\\ [4pt]
			
			\multicolumn{1}{c}{} & \multicolumn{1}{c}{signal cavity frequency} & \multicolumn{1}{c}{} & \multicolumn{1}{c}{} & \multicolumn{7}{c}{\multirow{2}{*}{6.8330}} & \multicolumn{1}{c}{} & \multicolumn{1}{c}{} & \multicolumn{7}{c}{\multirow{2}{*}{2.8691}} &  \multicolumn{1}{c}{}\\
			\multicolumn{1}{c}{} & \multicolumn{1}{c}{$\omega_{s}/2\pi$\,(GHz)} & \multicolumn{1}{c}{} & \multicolumn{1}{c}{} & \multicolumn{7}{c}{} & \multicolumn{1}{c}{} & \multicolumn{1}{c}{} & \multicolumn{7}{c}{} & \multicolumn{1}{c}{}\\ [4pt]
			
			\multicolumn{1}{c}{} & \multicolumn{1}{c}{signal cavity decay rate} & \multicolumn{1}{c}{} & \multicolumn{1}{c}{} & \multicolumn{7}{c}{\multirow{2}{*}{7}} & \multicolumn{1}{c}{} & \multicolumn{1}{c}{} & \multicolumn{7}{c}{\multirow{2}{*}{3}} &  \multicolumn{1}{c}{}\\
			\multicolumn{1}{c}{} & \multicolumn{1}{c}{$\kappa_{s}/2\pi$\,(kHz)} & \multicolumn{1}{c}{} & \multicolumn{1}{c}{} & \multicolumn{7}{c}{} & \multicolumn{1}{c}{} & \multicolumn{1}{c}{} & \multicolumn{7}{c}{} &  \multicolumn{1}{c}{}\\ [4pt]
			
			\multicolumn{1}{c}{} & \multicolumn{1}{c}{parameter coupling strength} & \multicolumn{1}{c}{} & \multicolumn{1}{c}{} & \multicolumn{7}{c}{\multirow{2}{*}{56.569}} & \multicolumn{1}{c}{} & \multicolumn{1}{c}{} & \multicolumn{7}{c}{\multirow{2}{*}{1.697$\times10^4$}} & \multicolumn{1}{c}{}\\
			\multicolumn{1}{c}{} & \multicolumn{1}{c}{$J/2\pi$\,(kHz)} & \multicolumn{1}{c}{} & \multicolumn{1}{c}{} & \multicolumn{7}{c}{} & \multicolumn{1}{c}{} & \multicolumn{1}{c}{} & \multicolumn{7}{c}{} & \multicolumn{1}{c}{}\\ [4pt]
			
			\multicolumn{1}{c}{} & \multicolumn{1}{c}{SQUID's pumping frequency} & \multicolumn{1}{c}{} & \multicolumn{1}{c}{} & \multicolumn{7}{c}{\multirow{2}{*}{10}} & \multicolumn{1}{c}{} & \multicolumn{1}{c}{} & \multicolumn{7}{c}{\multirow{2}{*}{2}} &  \multicolumn{1}{c}{}\\
			\multicolumn{1}{c}{} & \multicolumn{1}{c}{$\omega_{\text{SQUID}}/2\pi$\,(GHz)} & \multicolumn{1}{c}{} & \multicolumn{1}{c}{} & \multicolumn{7}{c}{} & \multicolumn{1}{c}{} & \multicolumn{1}{c}{} & \multicolumn{7}{c}{} &  \multicolumn{1}{c}{}\\ [4pt]
			
			\multicolumn{1}{c}{} & \multicolumn{1}{c}{pump cavity frequency} & \multicolumn{1}{c}{} & \multicolumn{1}{c}{} & \multicolumn{7}{c}{\multirow{2}{*}{3.6660}} & \multicolumn{1}{c}{} & \multicolumn{1}{c}{} & \multicolumn{7}{c}{\multirow{2}{*}{3.7389}} &  \multicolumn{1}{c}{} \\
			\multicolumn{1}{c}{} & \multicolumn{1}{c}{$\omega_{p}/2\pi$\,(GHz)} & \multicolumn{1}{c}{} & \multicolumn{1}{c}{} & \multicolumn{7}{c}{} & \multicolumn{1}{c}{} & \multicolumn{1}{c}{} & \multicolumn{7}{c}{} &  \multicolumn{1}{c}{}\\
			
			\cmidrule(r){4-12} \cmidrule{13-21}
			
			\multicolumn{1}{c}{} & \multicolumn{1}{c}{external field} & \multicolumn{1}{c}{} & \multicolumn{1}{c}{} & \multicolumn{1}{c}{driving} & \multicolumn{1}{c}{} &  \multicolumn{1}{c}{} & \multicolumn{1}{c}{driving} & \multicolumn{1}{c}{} & & \multicolumn{1}{c}{no driving} & \multicolumn{1}{c}{} & \multicolumn{1}{c}{} & \multicolumn{1}{c}{driving} & \multicolumn{1}{c}{} & \multicolumn{1}{c}{} & \multicolumn{1}{c}{driving} & \multicolumn{1}{c}{} & \multicolumn{1}{c}{} & \multicolumn{1}{c}{no driving} &  \multicolumn{1}{c}{} \\
			
			\multicolumn{1}{c}{} & \multicolumn{2}{c}{} & \multicolumn{6}{c}{} & \multicolumn{3}{c}{} & \multicolumn{6}{c}{} & \multicolumn{3}{c}{} \\[-8pt]
			
			\multicolumn{1}{c}{} & \multicolumn{1}{c}{driving amplitude} & \multicolumn{1}{c}{} & \multicolumn{1}{c}{} & \multicolumn{1}{c}{\multirow{2}{*}{10}} & \multicolumn{1}{c}{} &  \multicolumn{1}{c}{} & \multicolumn{1}{c}{\multirow{2}{*}{10}} & \multicolumn{1}{c}{} & \multicolumn{1}{c}{} & \multicolumn{1}{c}{\multirow{2}{*}{---}} & \multicolumn{1}{c}{} & \multicolumn{1}{c}{} & \multicolumn{1}{c}{\multirow{2}{*}{10}} & \multicolumn{1}{c}{} &  \multicolumn{1}{c}{} & \multicolumn{1}{c}{\multirow{2}{*}{10}} & \multicolumn{1}{c}{} & \multicolumn{1}{c}{} & \multicolumn{1}{c}{\multirow{2}{*}{---}} &  \multicolumn{1}{c}{}\\
			\multicolumn{1}{c}{} & \multicolumn{1}{c}{$\Omega/2\pi$\,(MHz)} & \multicolumn{1}{c}{} & \multicolumn{6}{c}{} & \multicolumn{3}{c}{} & \multicolumn{6}{c}{} & \multicolumn{3}{c}{} \\[4pt]
			
			\multicolumn{1}{c}{} & \multicolumn{1}{c}{pump cavity decay rate} & \multicolumn{1}{c}{} & \multicolumn{1}{c}{} & \multicolumn{1}{c}{\multirow{2}{*}{10}} & \multicolumn{1}{c}{} &  \multicolumn{1}{c}{} & \multicolumn{1}{c}{\multirow{2}{*}{10}} & \multicolumn{1}{c}{} & \multicolumn{1}{c}{} & \multicolumn{1}{c}{\multirow{2}{*}{$0.003$}} & \multicolumn{1}{c}{} & \multicolumn{1}{c}{} & \multicolumn{1}{c}{\multirow{2}{*}{10}} & \multicolumn{1}{c}{} &  \multicolumn{1}{c}{} & \multicolumn{1}{c}{\multirow{2}{*}{10}} & \multicolumn{1}{c}{} & \multicolumn{1}{c}{} & \multicolumn{1}{c}{\multirow{2}{*}{$0.003$}} &  \multicolumn{1}{c}{}\\
			\multicolumn{1}{c}{} & \multicolumn{1}{c}{$\kappa_{p}/2\pi$\,(MHz)} & \multicolumn{1}{c}{} & \multicolumn{6}{c}{} & \multicolumn{3}{c}{} & \multicolumn{6}{c}{} & \multicolumn{3}{c}{}\\[4pt]
			
			\multicolumn{1}{c}{} & \multicolumn{2}{c}{} & \multicolumn{3}{c}{} & \multicolumn{3}{c}{} & \multicolumn{3}{c}{} & \multicolumn{3}{c}{} & \multicolumn{3}{c}{} & \multicolumn{3}{c}{} \\[-8pt]
			
			\multicolumn{1}{c}{} & \multicolumn{1}{c}{amplitude of the initial} & \multicolumn{1}{c}{} & \multicolumn{1}{c}{} & \multicolumn{1}{c}{\multirow{2}{*}{$0$}} & \multicolumn{1}{c}{} & \multicolumn{1}{c}{} & \multicolumn{1}{c}{\multirow{2}{*}{$i$}} & \multicolumn{1}{c}{} & \multicolumn{1}{c}{} & \multicolumn{1}{c}{\multirow{2}{*}{$i$}} & \multicolumn{1}{c}{} & \multicolumn{1}{c}{} & \multicolumn{1}{c}{\multirow{2}{*}{$0$}} & \multicolumn{1}{c}{} & \multicolumn{1}{c}{} & \multicolumn{1}{c}{\multirow{2}{*}{$i$}} & \multicolumn{1}{c}{} & \multicolumn{1}{c}{} & \multicolumn{1}{c}{\multirow{2}{*}{$i$}} &  \multicolumn{1}{c}{}\\
			\multicolumn{1}{c}{} & \multicolumn{1}{c}{coherent state $\alpha$} & \multicolumn{1}{c}{} & \multicolumn{3}{c}{} & \multicolumn{3}{c}{} & \multicolumn{3}{c}{} & \multicolumn{3}{c}{} & \multicolumn{3}{c}{} & \multicolumn{3}{c}{}\\[4pt]	
			
			\multicolumn{3}{c}{} & \multicolumn{3}{c}{} & \multicolumn{3}{c}{} & \multicolumn{3}{c}{} & \multicolumn{3}{c}{} & \multicolumn{3}{c}{} & \multicolumn{3}{c}{} \\[-8pt]
			
			\multicolumn{1}{c}{} & \multicolumn{1}{c}{minimum of the spin squeezing} & \multicolumn{1}{c}{} & \multicolumn{1}{c}{} & \multicolumn{1}{c}{\multirow{2}{*}{$-15.13$}} & \multicolumn{1}{c}{} & \multicolumn{1}{c}{} & \multicolumn{1}{c}{\multirow{2}{*}{$-14.91$}} & \multicolumn{1}{c}{} & \multicolumn{1}{c}{} & \multicolumn{1}{c}{\multirow{2}{*}{$-2.34$}} & \multicolumn{1}{c}{} &  \multicolumn{1}{c}{} & \multicolumn{1}{c}{\multirow{2}{*}{$-13.58$}} & \multicolumn{1}{c}{} &  \multicolumn{1}{c}{} & \multicolumn{1}{c}{\multirow{2}{*}{$-13.58$}} & \multicolumn{1}{c}{} & \multicolumn{1}{c}{} & \multicolumn{1}{c}{\multirow{2}{*}{$-9.51$}} &  \multicolumn{1}{c}{}\\
			\multicolumn{1}{c}{} & \multicolumn{1}{c}{parameter $\xi^2_{\mathcal{R},\min}$\,(dB)\tnote{1}} & \multicolumn{1}{c}{} & \multicolumn{3}{c}{} & \multicolumn{3}{c}{} & \multicolumn{3}{c}{} & \multicolumn{3}{c}{} & \multicolumn{3}{c}{} & \multicolumn{3}{c}{}\\[4pt]
			
			\hline\\[-9pt]
			\hline
			
		\end{tabular}
		\begin{tablenotes}   
			\footnotesize              
			\item[1] Parameter $\xi^2_{\mathcal{R},\min}$ differs from the parameter $\xi^2_{R,\min}$ in Sec.~\ref{SEC3}. The details are shown in \ref{APPb}.
		\end{tablenotes}     
	\end{threeparttable}
\end{table*} 
	In order to demonstrate further the performance of the protocol, we discuss the experimental feasibility by combining the theoretical model in the protocol with the current experiments.
	
\begin{figure}
		\subfigure{\includegraphics[scale=0.60]{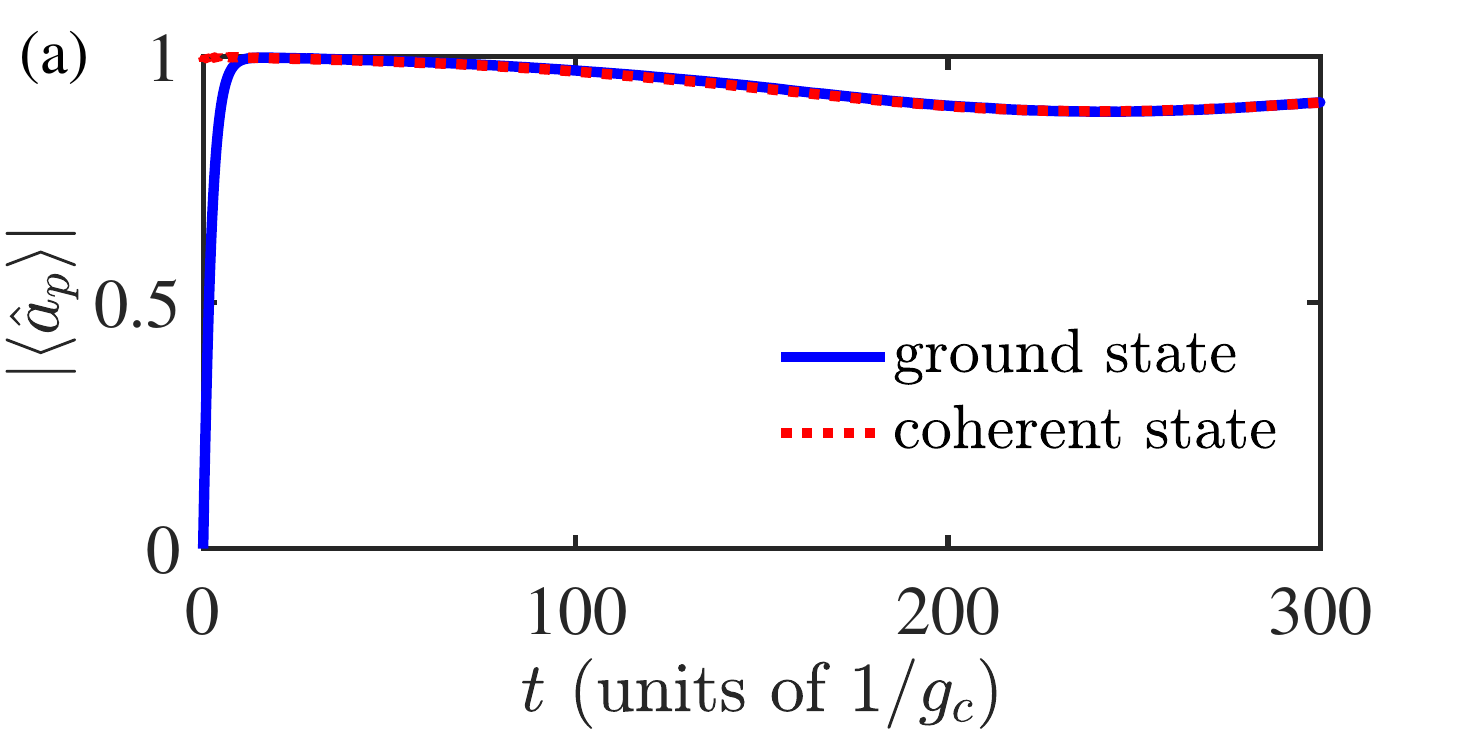}\label{fig14a}}
		\subfigure{\includegraphics[scale=0.60]{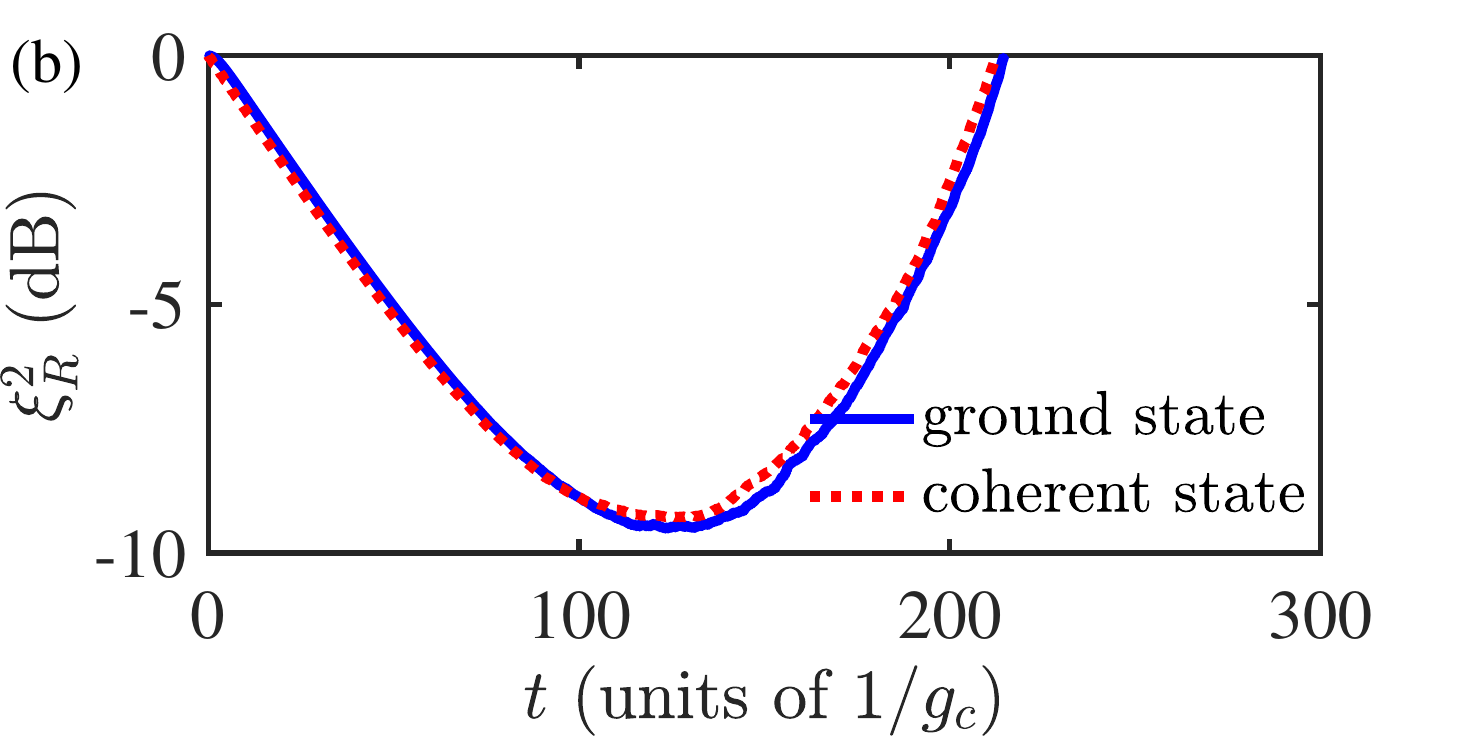}\label{fig14b}}
		\caption[Fig1]{Time evolution of (a) the parameter $\vert\langle\hat{a}_p\rangle\vert$, and (b) the spin squeezing parameter $\xi^2_R$ for the pump cavity initiated in the ground state (the blue solid curve) and the quasi-steady state (the red dotted curve), given by the full master equation in Eq.~(\ref{sec2eq22}). All other parameters are ${N=50}$, ${\delta_{s}=15g_c}$,  $\Omega=\kappa_{p}$, $J=\sqrt{2}g_c$, and $\kappa_{s}=\gamma_{c}=\gamma_{s}=0$.} 
		\label{fig14}
	\end{figure}

	Here, we consider that a setup consisting of two coplanar waveguide resonators (CPWRs), a superconducting quantum interference device (SQUID), and a rubidium (Rb) atomic ensemble, as shown in Fig.\,\ref{fig12}. The SQUID is used to mediate the parametric conversion between a single photon of the pump cavity (CPWR I) and a pair of photons of the signal cavity (CPWR II). In other words, the SQUID constructs the parametric coupling between two cavities. The pumping frequency of the SQUID satisfies ${\omega_{\text{SQUID}}=2\omega_{s}-\omega_{p}}$. The SQUID has been designed and realized in many experiments, like the asymmetrically threaded SQUID (ATS) \cite{Lescanne2020} and the rf-SQUID \cite{Vrajitoarea2019}. The strength of the parametric coupling constructed by the SQUID has been reported with a range of ${J/2\pi=0.1\sim17.7}$ MHz \cite{Leghtas2015,Chang2020,Lescanne2020,Vrajitoarea2019}. Meanwhile, we take the atomic clock states ${\{5S^{1/2},\,F=1,\,m_F=-1\}:=\vert1,\,-1\rangle}$ and ${\{5S^{1/2},\,F=2,\,m_F=1\}:=\vert2,\,1\rangle}$ as the information carrier \cite{Hattermann2017,PhysRevLett.103.043603}. To date, coupling these two states to the CPWR has been experimentally implemented with an additional rf field \cite{Hattermann2017}. However, the reported coupling strength is too weak to implement efficient quantum coherent operations \cite{PhysRevResearch.4.013207}. Fortunately, several approaches can be used to increase the coupling strength. Such approaches include decreasing the space between the atomic ensemble and the CPWR \cite{Hattermann2017}, and introducing novel CPWRs of being able to provide some strong magnetic fields \cite{PhysRevApplied.11.064053}. Thus, here, we take $g_c/2\pi=40$ kHz with $N=10^6$ \cite{PhysRevLett.103.043603}. Note that, both the relaxation time $T_1$ and the coherent time $T_2$ of the Rb atomic ensemble are of the order sec \cite{Bernon2013}. This indicates that one can ignore the effect of the atomic spontaneous emission and the collective dephasing on the dynamics of the system. In addition, for the CPWRs, the quality factor over $10^6$ has been realized experimentally \cite{doi:10.1063/1.4919761,Mirhosseini2018,RevModPhys.93.025005}. Thus, we take ${\kappa_{p}\sim10^{-6}\omega_{p}}$ and ${\kappa_{s}\sim10^{-6}\omega_{s}}$. According to these parameters mentioned above, we list a group of feasible parameters and then estimate the minimum of the spin squeezing parameter in Table~\ref{table1}. Note that, for the case with driving (discussed in Sec.~\ref{SEC3a}), it is not necessary that the initial state of the pump cavity is the quasi-steady coherent state. As shown in Fig.~\ref{fig14}, the generation of spin squeezing is also able to be achieved when the pump cavity is in the ground state initially. Meanwhile, from Fig.~\ref{fig14}, one can find that the generations of spin squeezing between these two cases with different initial states of the pump cavity are almost the same. The reason of this similarity is that the time cost of constructing the quasi-steady coherent state from the ground state in the pump cavity is too short to change the generation of spin squeezing significantly.
	
	Furthermore, the protocol can also be implemented with ensembles of other particles, such as nitrogen-vacancy (NV) centers in diamond \cite{PhysRevLett.107.060502,PhysRevLett.105.140502,PhysRevA.85.012333,PhysRevLett.118.140502,Putz2014,PhysRevB.82.201201,Cheng:13,Han:17,PhysRevA.100.052332,Cheng2019,Cheng2016}. In this case, a feasible setup is the same as the one shown in Fig.~\ref{fig12}, but with replacing the atomic ensemble by the NV centers. We encode the internal states of the NV centers, ${\vert m_s=0\rangle}$ and ${\vert m_s=-1\rangle}$, as the information carrier \cite{Putz2014}. The coupling strength between a single NV center and a single CPWR photon $g/2\pi$ exceeds $12\,\text{Hz}$ \cite{PhysRevLett.107.060502,Putz2014}. To date, for a typical ensemble of NV centers, the relaxation time of ${T_1\sim40\,\text{s}}$ (${\gamma_{s}\ll1\,\text{Hz}}$) have been demonstrated experimentally \cite{PhysRevLett.107.060502} and, with the spin-echo sequences, the coherent time of ${T_2>600\,\mu\text{s}}$ (${\gamma_{c}<0.26\,\text{kHz}}$) has also been reached \cite{PhysRevB.82.201201,QinChenWangMiranowiczNori+2020+4853+4868}. Thus, we also give a group of feasible parameters and the minimum of the spin squeezing parameter in Table~\ref{table1}.

	\section{CONCLUSIONS}\label{SEC5}
	In this paper, we have proposed a protocol to generate spin squeezing in atomic ensembles via a fully quantum degenerate parametric amplifier. By adjusting the parameters, an effective cavity-induced TAT-like interaction can be achieved. The strength of the generated spin squeezing is determined by some properties of the pump cavity, such as the initial state, the driving strength, and the cavity decay. We mainly discuss the generated spin squeezing in two cases of the pump cavity. For the first case, the pump cavity is initially in the quasi-steady coherent state by a driving field and the cavity decay. Meanwhile, for the second case, the pump cavity is initiated in an arbitrary coherent state and there is no pump cavity driving.
	
	In the first case, theoretical analyses and numerical simulations show that the present protocol can generate a strong spin squeezing whose strength is even comparable to that of the TAT model. The reason is that, for a fixed $d_0$ and an extremely large ratio of $\kappa_{p}$ to $g_\text{eff}$, the effect of the cavity-induced TAT-like interaction is equivalent to applying the TAT interaction to the atomic ensemble. Meanwhile, according to numerical simulations, the present protocol is robust to the atomic spontaneous emission and the signal cavity decay.
		 
	In the second case, according to numerical simulations, the present protocol can generate an observable spin squeezing. A properly large amplitude of the initial coherent state can accelerate the generation of spin squeezing. Meanwhile, the present protocol is also robust to the collective dephasing of the ensemble, in addition to the atomic spontaneous emission and the signal cavity decay. It is worth noting that the pump cavity decay is able to prolong the duration of squeezing significantly.
	
	After combining the experimental results, we show that the present protocol is feasible experimentally. A group of realistic parameters, as shown in Table \ref{table1}, has been given to predict some experimentally feasible results. Moreover, the present protocol can be extended to generate spin squeezing in various ensembles, like ensembles of NV centers. We hope that the present protocol provides a novel approach for generating spin squeezing in photon-spin coupling systems.
	
	\begin{acknowledgments}
		This work was supported by the National Natural Science Foundation of China under Grants No. 11575045, No. 11874114, and No. 11674060, the Natural Science Funds for Distinguished Young Scholar of Fujian Province under Grant No. 2020J06011 and Project from Fuzhou University under Grant No. JG202001-2. Y.-H.C. was supported by the Japan Society for the Promotion of Science (JSPS) KAKENHI Grant No. JP19F19028. W.Q. is supported in part by the Incentive Research Project of RIKEN.	
	\end{acknowledgments}
	 
	\begin{appendices} 
	\renewcommand\thesection{Appendix~\Alph{section}}
	\section{THE APPROACH FOR COMPENSATING THE TERM $(g^2/\delta_{\texttt{S}})\hat{S}_+\hat{S}_-$ COMPLETELY}\label{aX}
	\setcounter{equation}{0}
	\renewcommand\theequation{A\arabic{equation}}
	In this appendix, we give the detailed derivation of the approach for compensating $(g^2/\delta_{\texttt{S}})\hat{S}_+\hat{S}_-$ completely. We first assume the information of spin squeezing is carrier by the levels of the atoms $\vert0\rangle_a\equiv\vert\downarrow\rangle_a$ and $\vert1\rangle_a\equiv\vert\uparrow\rangle_a$. Here, we introduce an optical cavity, a laser pulse, and an auxiliary level of the atoms $\vert 2\rangle_a$ \cite{PhysRevLett.113.203601,Lauk_2020,PhysRevA.103.023706}. The optical cavity (pulse) is far-off resonant with the transition $\vert 0\rangle_a\leftrightarrow\vert 2\rangle_a$ ($\vert 1\rangle_a\leftrightarrow\vert 2\rangle_a$) with the coupling strength $g_{d}$ ($\Omega_{o}$) and the detuning $\Delta_{d}$ ($\Delta_{o}$). In the interaction picture, the Hamiltonian described these interactions is given as
		\begin{eqnarray}\label{xA1}
			\hat{H}_o&=&\Delta_{d}\hat{d}^\dagger\hat{d}+g_c\sum_{k=1}^{N}(\vert 2\rangle^k_a{^k_a\langle0\vert}\hat{c}+\vert 0\rangle^k_a{^k_a\langle2\vert}\hat{c}^\dagger)\cr\cr
			&&+\Omega_{o}\sum_{k=1}^{N}(\vert 2\rangle^k_a{^k_a\langle1\vert}e^{-i\Delta_{o}t}+\vert 1\rangle^k_a{^k_a\langle2\vert}e^{i\Delta_{o}t}), 
		\end{eqnarray}
		where $\hat{d}^\dagger$ and $\hat{d}$ are the creation operator and the annihilation operator of the optical cavity mode, respectively. $\vert\varepsilon\rangle^k_a\,(\varepsilon=0,1,2\ \text{and}\ k=1,2,\dots,N)$ represents that the $k$th atom is in $\vert\varepsilon\rangle_a$. When the conditions $\Delta_{o}-\Delta_{d}=\Delta$ and $\{\Delta_{d},\Delta_{o}\}\gg\{g_d,\Omega_{o},\Delta\}$ are satisfied, the auxiliary level $\vert 2\rangle_a$ can be eliminated adiabatically and then the effective Hamiltonian of the system is
		\begin{eqnarray}\label{xA2}
			\hat{H}^{(1)}_{o,\text{eff}}&=&\dfrac{g_d\Omega_{o}}{\Delta'}\sum_{k=1}^{N}(\vert1\rangle^k_a{^k_a\langle0\vert}\hat{d}e^{i\Delta t}+\vert 0\rangle^k_a{^k_a\langle1\vert}\hat{d}^\dagger e^{-i\Delta t})\cr\cr
			&&+\dfrac{\Omega_{o}^2}{\Delta_o}\sum_{k=1}^{N}\vert 1\rangle^k_a{^k_a\langle1\vert}+\dfrac{g_{d}^2}{\Delta_d}\sum_{k=1}^{N}\vert 0\rangle^k_a{^k_a\langle0\vert}\hat{c}^\dagger\hat{c},
		\end{eqnarray}
		where $\Delta'=2\Delta_{d}\Delta_{o}/(\Delta_{o}+\Delta_{d})$. To adiabatically eliminate the optical cavity mode, we set $\Delta\gg g_d\Omega_{o}/\Delta'$. Then, the above effective Hamiltonian in Eq.~(\ref{xA2}) is reduced to
		\begin{eqnarray}\label{xA3}
			\hat{H}^{(2)}_{o,\text{eff}}&=&\dfrac{g^2_d\Omega^2_{o}}{\Delta^{\prime2}\Delta}\hat{S}_+\hat{S}_-+\dfrac{\Omega_{o}^2}{\Delta_o}\hat{S}_z,
		\end{eqnarray}
		where $\hat{S}_+\hat{S}_-=\sum_{k=1}^{N}\sum_{k'=1}^{N}(\vert1\rangle^k_a{^k_a\langle0\vert})(\vert 0\rangle^{k'}_a{^{k'}_a\langle1\vert})$ and $\hat{S}_z=1/2\sum_{k=1}^{N}(\vert 1\rangle^k_a{^k_a\langle1\vert}-\sum_{k=1}^{N}\vert 0\rangle^k_a{^k_a\langle0\vert})$. This means that, when one takes
		\begin{eqnarray}\label{xA4}
			\dfrac{g^2_d\Omega^2_{o}}{\Delta^{\prime2}\Delta}+\dfrac{g^2}{\delta_{\texttt{S}}}=\delta_{q}+\dfrac{\Omega_{o}^2}{\Delta_o}=0,
		\end{eqnarray}
		the effect of the undesired term $(g^2/\delta_{\texttt{S}})\hat{S}_+\hat{S}_-$ is compensated completely.

	\section{THE ADIABATIC ELIMINATION OF THE SIGNAL CAVITY}\label{APPa}
	\setcounter{equation}{0}
	\renewcommand\theequation{B\arabic{equation}}
	Due to few photons and strong cavity loss, the signal cavity can be considered as an ambience and can be adiabatically eliminated.	Here, we give a detailed derivation on the adiabatic elimination of the signal cavity. According to the Hamiltonian in Eq.~(\ref{sec2eq2}), one can obtain the interactions which excite signal cavity photons
	\begin{equation}\label{appa1}
		\hat{V}_1=J\hat{a}_p\hat{a}^{\dagger2}_s,\ \hat{V}_2=g\hat{S}_-\hat{a}^{\dagger}_s,
	\end{equation}
	and the free Hamiltonian of signal cavity photons
	\begin{equation}\label{appa2}
		\hat{H}_e=\delta_{s}\hat{a}^\dagger_s\hat{a}_s.
	\end{equation}
	We introduce a Lindblad operator $L_{s}=\sqrt{\gamma_{s}}\hat{a}_s$ which satisfies $\mathcal{L}(L_s)\rho=\gamma_{s}\mathcal{L}(\hat{a}_s)\rho$. According to the work in Ref.~\cite{PhysRevA.85.032111}, the Lindblad operators of the effective master equation can be described as
	\begin{equation}\label{appa3}
		\hat{L}_h=\hat{L}_s\left(\hat{H}_e-\dfrac{i}{2}\hat{L}^\dagger_s\hat{L}_s\right)^{-1}\hat{V}_h,
	\end{equation}
	where, $h=1,2$. Then, one would obtain 
	\begin{eqnarray}\label{appa4}
		\hat{L}_1&=&\dfrac{2\sqrt{\gamma_{s}}J}{2\delta_{s}-i\gamma_s}\hat{a}_p\hat{a}^\dagger_s,\cr\cr
		\hat{L}_2&=&\dfrac{2\sqrt{\gamma_{s}}g}{2\delta_{s}-i\gamma_s}\hat{S}_-.
	\end{eqnarray}
	This means that the Lindblad superoperator which describes the signal cavity decay $\gamma_{s}\mathcal{L}(\hat{a}_s)\rho$ can be replaced by $J^2p_\gamma\mathcal{L}(\hat{a}_p\hat{a}^\dagger_s)\rho+g^2p_\gamma\mathcal{L}(\hat{S}_-)\rho$, where $p_\gamma=4\gamma_s/(4\delta_{s}^2+\gamma_s^2)$. Moreover, since the signal cavity remains in vacuum during the evolution of the system effectively, the term $J^2p_\gamma\mathcal{L}(\hat{a}_p\hat{a}^\dagger_s)\rho$ is able to reduce to $J^2p_\gamma\mathcal{L}(\hat{a}_p)\rho$. Therefore, the effect of the signal cavity decay $\gamma_{s}\mathcal{L}(\hat{a}_s)\rho$ is equivalent to introducing an extra pump cavity decay $J^2p_\gamma\mathcal{L}(\hat{a}_p)\rho$ and an extra atomic collective spontaneous emission $g^2p_\gamma\mathcal{L}(\hat{S}_-)\rho$. The physics mechanics is that, when the signal cavity is eliminated adiabatically, the pump cavity and the ensemble can be considered to be coupled to a new vacuum bath (i.e., the vacuum signal cavity).
	
	\section{THE MINIMUM OF THE SPIN SQUEEZING PARAMETER IN TABLE~\ref{table1}}\label{APPb}
	\setcounter{equation}{0}
	\renewcommand\theequation{C\arabic{equation}}
	Note that, in typical ensembles, the number of particles is of the order of millions (e.g., Rb atoms) or trillions (e.g., NV centers). This means that it is extremely difficult to exactly estimate the full dynamics of the ensemble. In the following, we simplify the dynamics of the system and then obtain the minimum of spin squeezing effectively. With the Holstein-Primakoff transformation (HPT) \cite{PhysRev.58.1098} and in the limit of ${N\rightarrow\infty}$, the collective spin operators can be transformed to the bosonic operators as
	\begin{equation}\label{sec4eq1}
		\hat{S}_-\rightarrow \sqrt{N}\hat{b},\,\hat{S}_z\rightarrow -N/2+\hat{b}^\dagger\hat{b},
	\end{equation}
	where $\hat{b}$ is the bosonic annihilation operator. Accordingly, the spin squeezing parameter in Eq.~(\ref{sec3eq1}) is rewritten as
	\begin{equation}\label{sec4eq2}
		\xi^2_\mathcal{R}=\frac{N^2}{(N-2\langle\hat{b}^\dagger\hat{b}\rangle)^2}(1+2\langle\hat{b}^\dagger\hat{b}\rangle-2\vert\langle\hat{b}^2\rangle\vert). 
	\end{equation}
	Meanwhile, from the effective master equation in Eq.~(\ref{sec2eq23}) and the HPT, we obtain the Heisenberg equation \cite{PhysRevA.101.053818}
	\begin{subequations}\label{sec4eq3}
		\begin{eqnarray}
			\partial_t\hat{a}_p&=&-iNg_\text{eff}\hat{b}^2-i\Omega^*/2-(\kappa_p+J^2p_\gamma)\hat{a}_p/2,\\ \cr
			\partial_t\hat{b}^2&=&-2iNg_\text{eff}\hat{a}_p(2\hat{b}^\dagger\hat{b}+1)-(\gamma_s+g_c^2p_\gamma+2\gamma_{c})\hat{b}^2,\cr&&\\ 
			\partial_t\hat{b}^\dagger\hat{b}&=&-2iNg_\text{eff}(\hat{a}_p\hat{b}^{\dagger2}-\hat{a}_p^\dagger\hat{b}^2)-(\gamma_s+g_c^2p)\hat{b}^\dagger\hat{b},
 		\end{eqnarray}
	\end{subequations}
	where $\partial_t=\partial/\partial t$. Applying the mean-field	approximation, the instantaneous evolution of the average values of the operators is described as 
	\begin{subequations}\label{sec4eq4}
		\begin{eqnarray}
			\partial_t\langle\hat{a}_p\rangle&=&-iNg_\text{eff}\langle\hat{b}^2\rangle-i\Omega^*/2-(\kappa_p+J^2p_\gamma)\langle\hat{a}_p\rangle/2,\cr &&\\ 
			\partial_t\langle\hat{b}^2\rangle&=&-2iNg_\text{eff}\langle\hat{a}_p\rangle(2\langle\hat{b}^\dagger\hat{b}\rangle{+}1)-(\gamma_s{+}g_c^2p_\gamma{+}2\gamma_{c})\langle\hat{b}^2\rangle,\cr&&\\ 
			\partial_t\langle\hat{b}^\dagger\hat{b}\rangle&=&-4iNg_\text{eff}\text{Im}(\langle\hat{a}_p\rangle\langle\hat{b}^{\dagger2}\rangle)-(\gamma_s+g_c^2p)\langle\hat{b}^\dagger\hat{b}\rangle.
		\end{eqnarray}
	\end{subequations}
	Then, the time evolutions of $\langle\hat{b}^2\rangle$ and $\langle\hat{b}^\dagger\hat{b}\rangle$ can be obtained by solving the coupled equations in Eq.~(\ref{sec4eq4}). After substituting $\langle\hat{b}^2\rangle$ and $\langle\hat{b}^\dagger\hat{b}\rangle$ into Eq.~(\ref{sec4eq2}), the evolution of the spin squeezing parameter $\xi^2_\mathcal{R}$ is also obtained and then the minimum of the spin squeezing parameter $\xi^2_{\mathcal{R},\min}$ can be achieved accordingly.	
	\end{appendices} 
	\bibliography{ref}
	 
\end{document}